\documentclass[preprintnumbers, superscriptaddress, showpacs, nofootinbib, amsfonts, amsmath, amssymb, aps, prd, notitlepage, floatfix, longbibliography]{revtex4-2}

\usepackage[utf8]{inputenc}
\usepackage{amsmath,amssymb,amsfonts,mathrsfs,bbold}
\usepackage{yfonts}
\usepackage{graphicx}
\usepackage[colorlinks=true,citecolor=blue,linkcolor=blue,urlcolor=blue]{hyperref}
\usepackage[dvipsnames]{xcolor}
\usepackage{lipsum}
\usepackage{slashed}
\usepackage[normalem]{ulem}
\usepackage{url}
\usepackage{physics}

\usepackage[top=1in, bottom=1in, left=1in, right=1in]{geometry}

\usepackage[utf8]{inputenc}

\usepackage{amssymb, bm, amsmath, graphicx, physymb}
\usepackage{enumerate}
\usepackage{amsfonts}
\usepackage{mathtools}
\usepackage{latexsym}
\usepackage{epstopdf}
\usepackage{subfigure}
\usepackage{tikz}
\usetikzlibrary{decorations.pathmorphing, decorations.markings, shapes.geometric, shapes.misc, calc, math}
\tikzstyle{gluon}=[decorate, decoration={coil,aspect=0.8, amplitude=1.5pt,  segment length=3pt}]
\usepackage{stackrel}
\usepackage[most]{tcolorbox}

\newcommand{\ben}{\begin{eqnarray*}}
\newcommand{\een}{\end{eqnarray*}}

\newcommand{\amu}{\alpha_\mu}

\allowdisplaybreaks

\begin{document}

\title{Adapting Public Exams for Introductory Physics: \\ A Mixed-Methods Investigation of Student Outcomes}

    \makeatletter  
\def\@fnsymbol#1{\ensuremath{\ifcase#1\or *\or \dagger\or \ddagger\or
   \mathsection\or \mathparagraph\or \|\or **\or \dagger\dagger
   \or \ddagger\ddagger \or \mathsection\mathsection \else\@ctrerr\fi}}
    \makeatother
    
\author{Binisha Marahatta }
\email[Email: ]{binisha@nmsu.edu}
\affiliation{Department of Physics, New Mexico State University, Las Cruces, New Mexico 88003, USA}
\author{Matthew~D.~Sievert}
\email[Email: ]{msievert@nmsu.edu}
\affiliation{Department of Physics, New Mexico State University, Las Cruces, New Mexico 88003, USA}


\begin{abstract}
Public Exams are an assessment system designed in the context of Biology Education Research to increase student engagement in the exam design process.  Students are shown incomplete versions of test questions ahead of time, with the format of the question revealed but some details withheld, to help them tailor their study habits.  Students are also given the chance to provide feedback on the questions, refining them for clarity before the actual exam.  
In this study, we examined the impact of Public Exams on the experiences and outcomes of $N = 42$ students in an introductory algebra-based physics course at New Mexico State University, a Hispanic-Serving Institution in the American Southwest.  Using a combination of online numerical surveys, open-ended written responses, and student grade data, we conducted a mixed-methods analysis of the impact of Public Exams on students in this class.  We also applied our method to re-analyze the publicly available data of Wiggins et al. 
Students reported various benefits of Public Exams, including reduced test anxiety and directing to core concepts, and we observed a significant positive correlation ($p = 3.1 \times 10^{-7}$) between the participation rate of students in editing public exam questions and their average exam grade.  But students also expressed dissatisfaction with the incomplete information provided on problems involving the application of procedural knowledge, and they had a less positive response rate than the control sample (comments not related to public exams).  In other qualitative responses, students also consistently highlighted the application of procedural knowledge as both the aspects most helpful to their learning and the most common recommendation for future improvement.
Our findings suggest that physics students at NMSU do benefit from Public Exams through reduced anxiety and directing to core concepts, but that the Public Exam system, as previously described, is not optimized for use in physics.  We attribute this to the particular emphasis in physics on applying procedural knowledge and suggest modifications of the Public Exam system to better tailor it to a physics context.

\end{abstract}


\date{\today}
\maketitle

\newpage

%
\section{Introduction}
%

%
\subsection{Assessment in Physics and Related Disciplines}
%

Physics, like many STEM disciplines, is experiencing a pedagogical renaissance powered by the dedicated subfield of physics education research (PER) and informed by discipline-based education research (DBER) in adjacent fields.  Physics plays a unique role as the fundamental basis for other science, technology and engineering disciplines\cite{mcdermott1999resource}; it requires a comprehensive fluency in the language of mathematics, and it attaches physical meaning to those concepts\cite{tuminaro2004understanding}.  Physics also emphasizes the high-order cognitive skills necessary to mathematically model a physical situation -- arguably the most important skill a physics student learns\cite{wells1995modeling,hestenes1987toward}. 
The predominant teaching strategy in physics is the traditional lecture format, but numerous studies have shown that this strategy does little to promote students' understanding of core concepts\cite{halloun1985initial, mcdermott1991millikan, hake1998interactive}.  This traditional method of lecture can be characterized as a passive transformation of information from the board to the students' notebooks, which frequently hinders real-time cognitive processing\cite{mazur1996science}.  In some cases, lecture-based teaching can constitute a barrier to meaningful learning of the topics \cite{tanel2008effects}.  
Physics students also confront other social factors specific to physics, such as cultural stereotypes of intellect and nerdiness, the belief that only a ``genius'' understands physics, and unspoken hierarchies that privilege theoretical branches of physics as more challenging and prestigious\cite{johansson2020negotiating}.  Leslie et al. \cite{leslie2015expectations} suggested that the culture of physics prioritizes innate ``brilliance'' over hard work, creating a structural barrier that disproportionately discourages women and underrepresented minorities from pursuing the field.  

Likewise, the traditional approaches to assessment in Physics have limited scope and may actively hinder students' development of higher-order cognition.  The simplest exam formats using multiple choice questions are a necessity for many large lecture classes, but these assessments incentivize students to memorize facts and do not test critical understanding\cite{arhin2015effect}.  More traditional textbook problem solving in physics involves open-ended calculations and word problems; these assessment methods have been shown to improve students' ability to manipulate formulas and equations (applying procedural knowledge) but have limited effects on conceptual understanding\cite{kim2002students}.  This limited cognitive depth applied to problem solving leads to the ``plug and chug'' approach in students; when combined with passive blackboard lectures, repetitive word problems cause students to resort to memorizing the equations and algorithms they previously copied\cite{mazur1996science}.  And just as assessment design can promote student learning \cite{Black2009}, it can also hinder student cognition and performance. In traditional exams, students also often report an inability to recall the information they knew beforehand: a ``freeze up'' or ``forget everything'' reaction\cite{slater1997impact}.  This may be a genuine reaction due to an intrinsic test anxiety or a learned anxiety from students' peers, or given as an excuse for underperformance.

Various innovative strategies in assessment have been explored that significantly improve students' learning outcomes and higher-order cognitive development.  These include both innovations in how the assessment is prepared as well as innovations in the nature of the assessment.  

Test Question Templates (TQTs) are a structural framework for designing multiple equivalent versions of problems whose general format is known to the students in advance\cite{crowther2020testing}.  By emphasizing the template of a question rather than a specific question itself, TQTs provide an open-ended tool for generating both formative and summative assessment questions.  As defined by Crowther, Wiggins, and Jenkins\cite{crowther2020testing}, a fully constructed TQT specifies the input and output to the question, a specific example, and a key or rubric which defines what full-credit and partial-credit responses look like.  TQTs thus represent one possible way of promoting alignment between formative and summative assessments and the systematic assessment of higher-order cognition.  The authors even note that the design of TQTs is intended to mimic the generalizability of formula-based problems in fields like physics and engineering.

Ref.~\cite{crowther2020testing} surveyed students who had experienced the use of TQTs, who recognized both the positive potential of Test Question Templates to help students focus their studies and to learn more deeply, as well as the potential dangers and confusion when introducing a new assessment methodology.  Crowther and Knight \cite{crowther2023using} illustrated how the TQTs can help instructors to perform ``self-audits'' to evaluate their coverage of learning objectives and alignment with assessments, thereby promoting evidence-based improvements to the curriculum.  Evans, Jenkins, and Crowther \cite{evans2023student} presented a mixed-methods study of student experiences and perceptions of TQTs.  They found that students demonstrated a high level of metacognitive engagement with TQTs (greater than 40\% voluntarily used TQTs to create their own additional examples), and that students consistently reported that TQTs helped them transfer their knowledge to high-stakes exams and that TQTs would also be helpful in other science courses.  Using similar methodology, Choi, Jenkins, and Crowther \cite{choi2025practice} showed that students overwhelmingly reported TQTs and their application in online discussion boards as beneficial to their learning.  Reported benefits to students include identifying gaps in knowledge, deepening knowledge, promoting interpersonal interactions, and understanding expectations for the exam.  Reported detriments included unreliable contributions from other students, inadequate instructor feedback, and a lack of engagement with the TQT assignment.  Interestingly, students reported these benefits at even higher levels after a few improvements were made to the methodology, including guaranteeing instructor feedback on the discussion boards and a commitment to integrate at least one student-generated problem on the exam.
More recently, Wiggins et al.\cite{wiggins2026exam} conducted a randomized controlled trial across multiple US community college science disciplines studying the impact of exam preview methods, including both TQT and Public Exams (discussed below).  They found that such preview methods significantly enhance student outcomes.  Students in the exam preview section showed significant learning gains (15\% higher gain score than control), enhancement in study habits, and a reduction of stress and negative experiences\cite{wiggins2026exam}. Furthermore, implementing these preview methods required only a modest faculty preparation cost (average of 62 additional minutes per week).

Similar work has focused on the use of multiple choice and clicker-based technology to lower physics-related anxiety and improve learning  \cite{molin2021toward, gil2023educational}.  Oral exams \cite{baghdadchi2024using} are another historically favored assessment method used in STEM disciplines.  Oral exams are associated with positive impacts for at-risk students in improving grades and enhancing motivation and confidence, but they are rarely implemented because of the difficulty in managing larger classes, maintaining grade consistency, and mitigating student anxiety or unfamiliarity with the exam format.  Other novel assessment frameworks include two-stage exams \cite{wieman2014physics}, in which students first complete the exam individually and then retake the same exam collaborating with peers, and problem-based learning \cite{moore2010evaluating, munoz2025problem, Tan2022}, in which students work in groups on a ``real-world'' problem and are assessed on knowledge, development, and accountability.

%
\subsection{Public Exams}
%

The main subject of this work is the innovative new assessment method called ``Public Exams,'' which was designed in the field of biology education\cite{wiggins2022public}.  Public Exams are intended to increase students' active engagement in the exam design process and are the result of attempts to re-design exams to engender trust and authentic engagement between students and instructors\cite{black1998assessment, driessen2022evaluating}.  In the Public Exam system, students are given partial versions of the test questions before the actual exam, with some information disclosed and some contents hidden.  Only a fraction of the exam questions are made public in this way.  Students are then given the opportunity to suggest edits and give feedback on the exam format and test questions, helping to refine them for better clarity before the exam, and thereafter the fully revealed versions of those questions are used on the actual Exam.  Public Exams are intended to increase engagement, reduce test anxiety, and help the students focus their study efforts on the most important topics (``directing to core concepts'').

\begin{figure}[t]
    \textbf{Question:} \qquad
    \begin{center}
    ``Three molecules are interacting with each other through the formation of three hydrogen bonds. \\ Draw three possible hydrogen bonds with dashed lines.  \\ Make the relevant partial charges clear.''
    \end{center}

    \vspace{0.25cm}

    \textbf{Public Exam:} \qquad ``The three molecules are water, alanine, and [withheld].''
    \vspace{0.5cm}
    \includegraphics[width=0.90\textwidth]{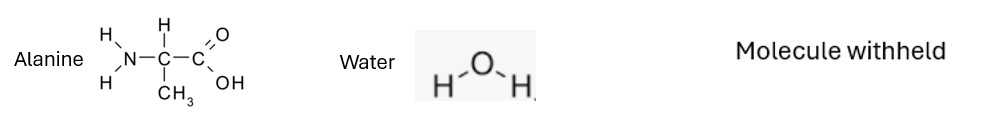}    
    
    \textbf{Actual Exam:} \qquad ``The three molecules are water, alanine, and acetic acid.''
    \includegraphics[width=0.90\textwidth]{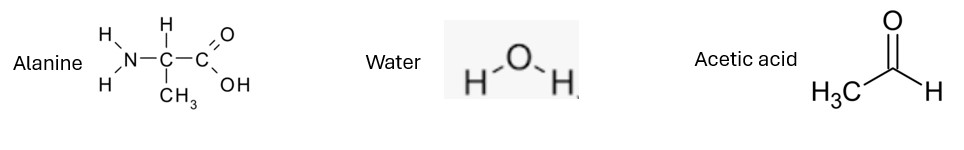}
    \caption{Example of a public question in Biology from Ref.~\cite{wiggins2022public}.  Top Panel: the question, Middle Panel: Information revealed on the Public Exam, Bottom Panel: Information presented to students on the actual exam.}
    \label{f:PublicExamsBio}
\end{figure}

An example of a public exam question in Biology, as implemented in Ref.~\cite{wiggins2022public} is shown in Fig.~\ref{f:PublicExamsBio}.  In the Public Exam which is released to the students ahead of time (middle panel), the third molecule is withheld; students then have an opportunity to provide edits on the Public Exam questions.  On the actual exam, the corresponding question is identical (bottom panel), with full information revealed. 
Public Exams clearly share some elements of Test Question Templates: they utilize the repetition of a particular problem type and encourage the student to imagine possible variations on the question.  Although the amount of information provided is far less than a full TQT \cite{crowther2020testing}, the Public Exam system further utilizes these simpler templates to revise draft versions of the exam.  The incorporation of corrections or even whole problems suggested by students further mirrors the improvements in methodology made in Ref.~\cite{choi2025practice} which showed greater benefits to students.

The impact of public exams on student experiences in a lower-division Biology course was studied by Wiggins et al. \cite{wiggins2022public} in a mixed-methods analysis of 292 students at an R1 institution and 32 students at a neighboring community college in the Pacific Northwest.  They collected qualitative data in the form of open-ended written responses and transcribed Zoom interviews with students as well as quantitative data in the form of exam and course grades.  They analyzed these data by conducting a thematic analysis of the written responses and interview transcripts and through a statistical analysis of grades based on factors like public versus traditional questions and engagement with the revision process.  In this paper, we study the application of the Public Exam system in a physics context and extend the analysis performed by Wiggins et al.

The authors of \cite{wiggins2022public} hypothesized that Public Exams would benefit students in four specific ways: (1) Language issues, (2) Directing to core concepts, (3) Deepening thought, and (4) Authentic engagement.  Their thematic analysis found evidence supporting these four themes, as well as two additional themes that emerged during the course of the research: (5) Anxiety or Confidence, and (6) Collaboration.  Their quantitative analysis found a very strong correlation ($p=0.000402$) between engagement (students who did versus did not suggest revisions to a public exam question) and overall course grades, as well as a $\sim 30\%$ better performance of students on public questions compared to traditional questions.  Wiggins et al. \cite{wiggins2022public} also noted evidence of both reduced anxiety and increased self-efficacy, although they did not separately study the two.

In this study, we present the results of implementing the Public Exam system in an NMSU physics course for the first time. The primary research questions of this study are (1) whether we see the same benefits of Public Exams in this study as previously reported by Wiggins et al., and (2) whether there are any differences in impact of the Public Exam system in a physics context versus a biology context.

The rest of this paper is organized as follows.  In Sec.~\ref{sec:Methods} we describe the methodology for our study, including population selection, the adaptation of Public Exams for use in an introductory physics course, and data collection.  In Sec.~\ref{sec:Results} we describe the results of our qualitative analysis, quantitative analysis, and mixed-methods analysis of the data.  Our analysis combines quantitative grade and survey data with qualitative data coded from open-ended student responses.  This includes an application of our methodology to the publicly available data from Wiggins et al. in Sec.~\ref{sec:Wiggins} for comparison.  In Sec.~\ref{sec:Discussion} we discuss the interpretation and implications of the results for the use of Public Exams in an introductory physics course.  Finally, in Sec.~\ref{sec:Concl} we suggest future modifications to better adapt the Public Exam system to a physics context and outline further studies to build upon these developments on innovations in physics assessment.

%

%
\section{Methodology}
\label{sec:Methods}
%

This study was conducted at New Mexico State University (NMSU), the public land-grant research university of New Mexico.  NMSU is a Carnegie R1 research institution and is a Hispanic-Serving Institution, as determined by criteria set by the US Department of Education: an enrollment of at least 25\% Hispanic students by full-time equivalence in the preceding year \cite{url4}.  
%
\begin{figure}[t]
    \centering
   \includegraphics[width=0.35\textwidth]{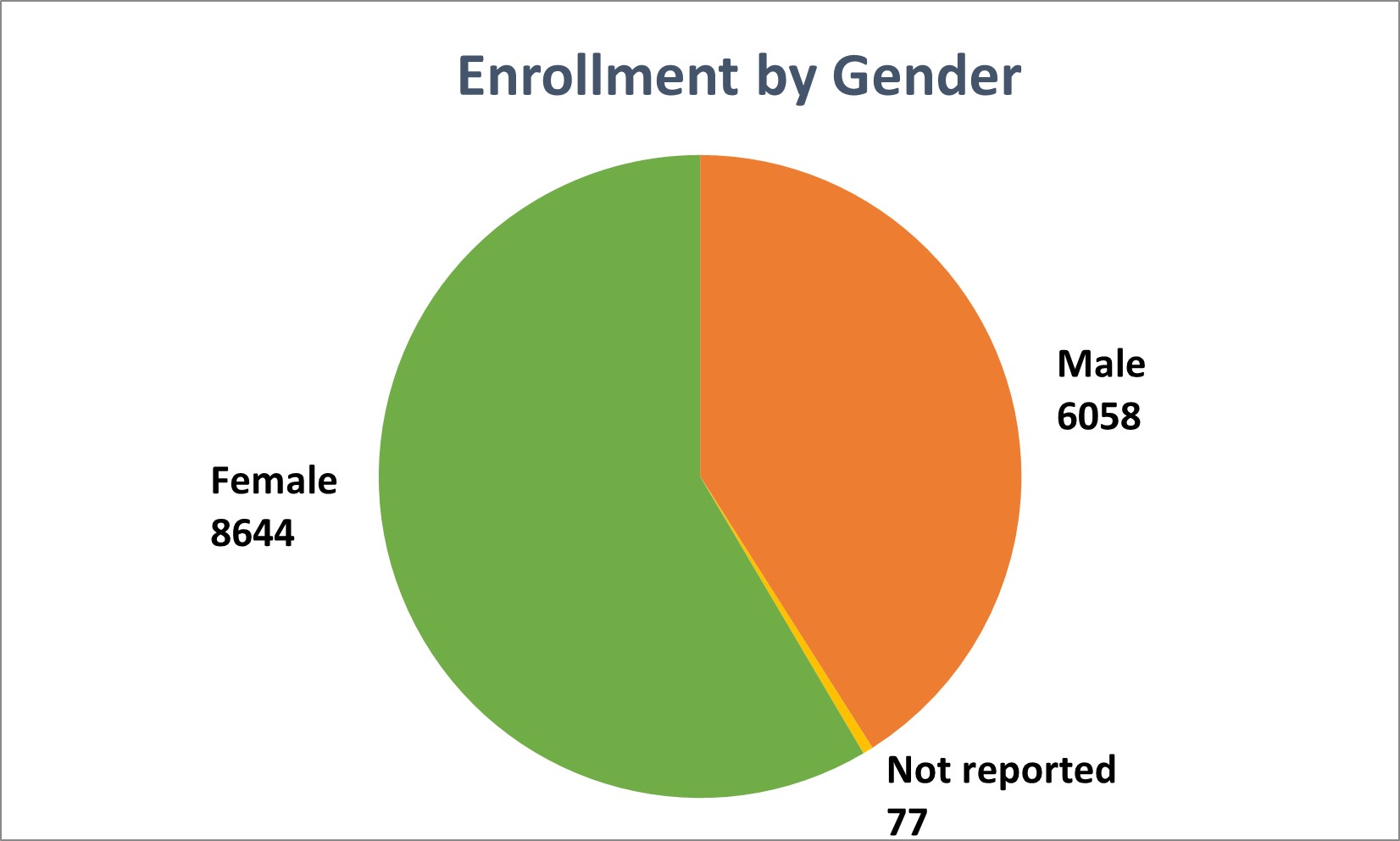}
    \includegraphics[width=0.55\textwidth]{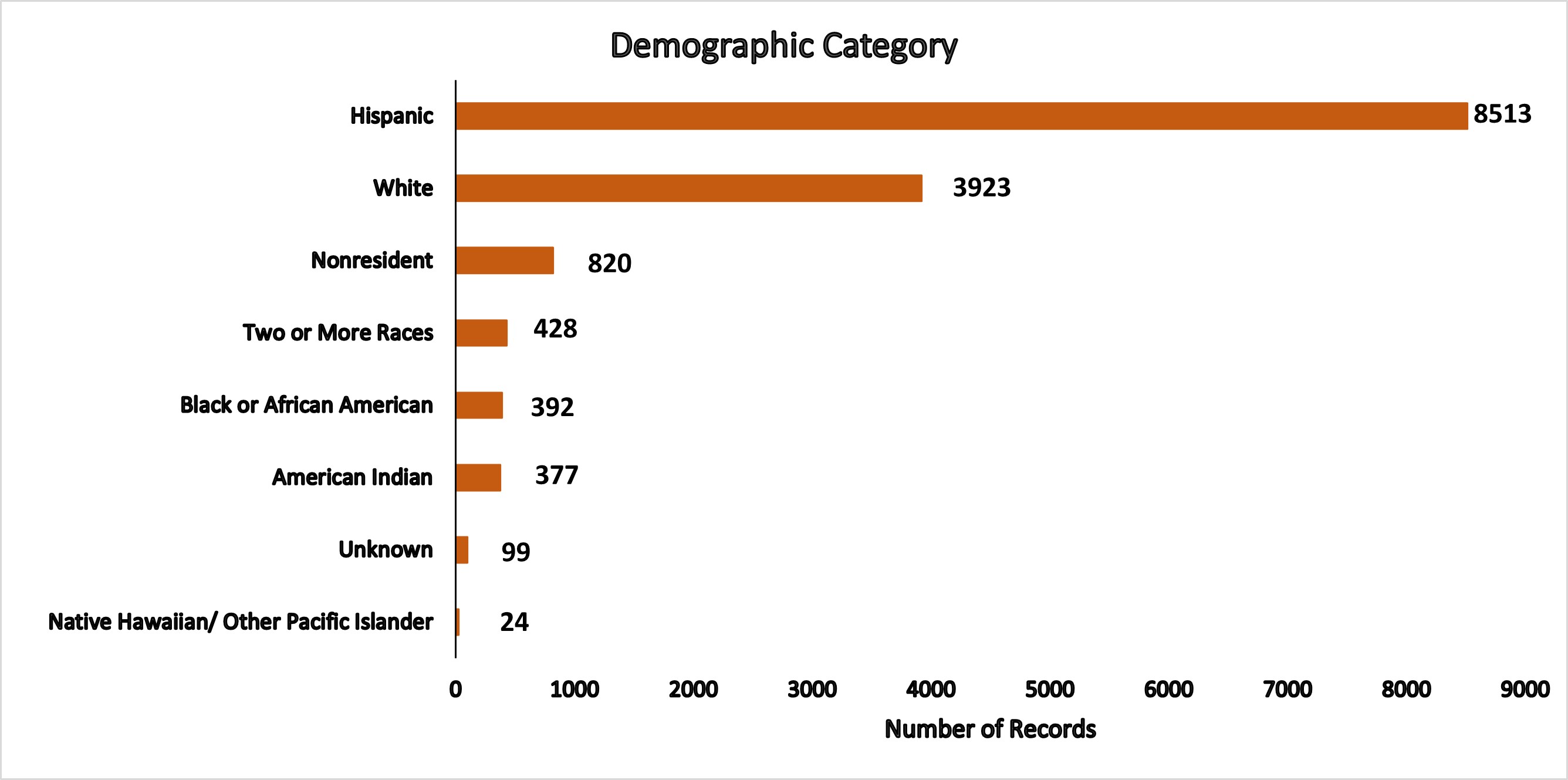}
    \caption{2023 NMSU student body by gender and {demographic category}.}
    \label{f:ethnicity}
\end{figure}
%
%
The data for this study was collected in 2023. The official demographics reported by NMSU's Office of Institutional Analysis \cite{url1} in Fall 2023 show that the main Las Cruces campus of NMSU consisted of 14,779 students, with 12,113 undergraduate students and 2,666 graduate students.  The enrollment of the students by gender and demographic category is shown in Fig.~\ref{f:ethnicity}.  By gender, the Las Cruces campus had 6058 male (41.0\%) and 8644 female (58.5\%) students, with 77 (0.5\%) not reported.  By ethnicity, the student body was majority Hispanic (57.6\%), followed by White/Caucasian (26.5\%).  Other reported demographic categories include Black or African American (2.7\%), American Indian (2.6\%), Asian (1.4\%), Native Hawaiian/Other Pacific Islander (0.2\%), and two or more races (2.9\%).  
Physics students at NMSU come from a wide variety of backgrounds and prior experience levels (especially in mathematics) and include nontraditional students, international students, and first-generation college students.  Students also experience a variety of barriers to learning that affect their outcomes, such as test anxiety, stereotype threat, and external demands on their time \cite{Aksian}, which affect the teaching and learning environment.

\begin{figure}[p]
    \textbf{Public Exam Version 1:} \qquad
    \begin{center}
    ``A circular loop of conducting wire is moving through a [UNIFORM OR NON-UNIFORM] magnetic field, \\ as shown in [FIGURE WITHHELD]. What is the nature of emf?''
    \end{center}
    \begin{enumerate}
    \item Non-Zero
    \item Zero emf
    \item The information on the question isn't sufficient to determine the answer. 
    \end{enumerate}

    \vspace{1cm}

    \textbf{Public Exam Version 2:} \qquad 
    \begin{center}
    \includegraphics[width=0.15\textwidth]{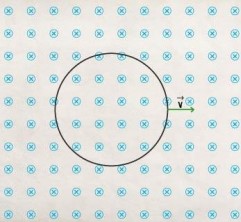} \\
    ``A circular loop of conducting wire is moving through a [UNIFORM OR NON-UNIFORM] magnetic field, \\ as shown in [FIGURE SHOWN]. What is the nature of emf?''
    \\ (UNIFORM MAGNETIC FIELD IS SHOWN HERE BUT FOR EXAM IT COULD BE NON-UNIFORM AS WELL) \\
    \end{center}
     \begin{enumerate}
    \item Non-Zero
    \item Zero emf
    \item The information on the question isn't sufficient to determine the answer. 
    \end{enumerate}
    \vspace{1cm}

    \textbf{Actual Exam:} \qquad 
    \begin{center}
    \includegraphics[width=0.15\textwidth]{PUBLIC_VERSION_2.jpeg} \\
    ``A circular loop of conducting wire is moving through a uniform magnetic field, \\ as shown in the figure. What is the nature of emf?''
    \end{center}
    \begin{enumerate}
    \item Non-Zero
    \item Zero emf
    \item The information on the question isn't sufficient to determine the answer. 
    \end{enumerate}

    \vspace{0.5cm}
    \caption{Example of a Public Exam question adapted for Physics in this study.  Top Panel: First version of the question, Middle Panel: Revised version incorporating student revisions, Bottom Panel: Question used on the actual Exam.}
    \label{f:PublicExamsPhys}
\end{figure}

Adapted to physics, Public Exams could benefit NMSU students in several ways, notably by reducing their test anxiety and by engendering trust and authentic engagement between students and instructors \cite{wiggins2022public}.  In this study, we implemented the Public Exam system in an introductory algebra-based physics course at NMSU.  There were two Midterm Exams and one Final Exam, with 50\% of the questions made public on the first Midterm, 57\% public questions on the second Midterm, and 60\% public questions on the Final Exam; the remaining non-public portions were administered in a traditional ``surprise" format with no prior knowledge of the questions.  The fraction of public questions increased over time in response to students' feedback (discussed in greater detail below).  Version 1 of the Public Exam was made available to the students a few days before the actual exam, and a second version (Version 2) which incorporated student revisions was given to students the day before the actual Exam. We note that the window of availability of the Public Exams to students was shorter in our study (a few days) than in Wiggins et al. \cite{wiggins2022public} (one week). The Version 2 questions are then used on the actual exam.  An example of this implementation is shown in Fig.~\ref{f:PublicExamsPhys}.

There were 42 participants in our research study, and we collected both qualitative and quantitative data from participants.  
The qualitative data includes open-ended handwritten responses to questions about the students' experiences on each of the three exams, while the quantitative analysis includes the grade data on the three exams (two midterms, one final exam). The open-ended written responses were collected after each midterm and immediately before the final exam. This study was approved and overseen by the NMSU Institutional Review Board (IRB), with 42 students consenting to the use of their grade data and 39 students consenting to the use of their qualitative data.  Individualized student data was anonymized by M. Sievert using a numeric ID before being provided to B. Marahatta for analysis, and the data were stored on a password-protected university drive.  There were no interactions between the participants and the research team other than securing participants' consent at the beginning of the semester.  A small extra credit incentive was offered to students for suggesting feedback on the public exams, as described below.

Another source of quantitative data was the Motivated Strategies for Learning Questionnaire (MSLQ) \cite{pintrich1991manual}.  The MSLQ is a calibrated survey instrument which measures both aspects of students' motivation and their learning strategies; we administered the portion related to student motivation using the online learning platform Canvas.  For each question, students rated themselves on the seven-point Likert scale from 1, representing ``not at all true of me'' to 7, representing ``very true of me.''  The MSLQ survey was given to all students as a part of the course (although only those consenting participants' data was used for this study), while suggesting feedback to a Public Exam question was optional but worth a small amount of extra credit (a $2\%$ bonus to the final grade for students who suggested feedback on all three public exams).

In the open-ended responses (collected after each midterm and immediately before the final exam), students responded to questions about their opinion regarding public exams, comparing the exams to each other, and whether using more public questions is beneficial to them.  Student responses were thematically coded by B. Marahatta for positive comments, negative comments, comments related to public exams, and comments related to the benefits specifically attributed to public exams (language issues, directing to core concepts, deepening thought, and engagement).  The research team discussed the results of the coding and reached consensus on any ambiguous cases.  Students generally described their likes and dislikes about the Public Exam system and its benefits to factors like anxiety; they also suggested improvements for the course and exam system in the future.

The coded open-ended responses are complemented by the quantitative data we collected regarding the engagement rate of the students in the Public Exam revision process and their performance on the actual exam. As described in detail in Sec.~\ref{sec:Results} below, we study the correlations of various factors including motivation, participation, and grade data.

%
\section{Results}
\label{sec:Results}
%

As mentioned above, the primary research questions of this study are (1) whether we see the same benefits of Public Exams reported by Wiggins et al (language issues, directing to core concepts, deepening thought, and authentic engagement) and (2) whether there are any differences in a physics context versus a biology context.  We present our results in four subsections: (1) Qualitative analysis, (2) Statistical analysis of qualitative data, (3) Other quantitative data and Mixed method analysis and (4) Re-analysis of Wiggins data.

%
\subsection{Qualitative Analysis}
%

%
\subsubsection{Comments Regarding Public Exams}
\label{sec:PEcomments}
%

\begin{figure}[t]
\centering
\includegraphics[width=0.4\linewidth]{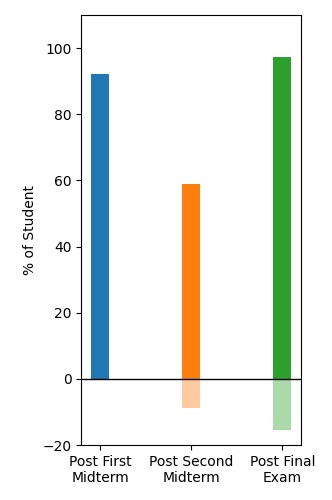}
 \caption{Open-ended responses of students for positive and negative comments for Public Exams where positive responses for different exams are shown in the axis above "0" and negative responses are shown in the axis below "0".  Percentage of students giving at least one positive comment and percentage of students giving at least one negative comment.}
\label{f:like/dislikes}
\end{figure}

In the open-ended written responses, there were several questions that were related to students' experiences and perceptions. These included: 
\begin{center}
\textit{
``What is your opinion about the public exam? Please compare and contrast it with the traditional exam.''}
\end{center}
\begin{center}
\textit{
``What is your opinion(s) about the mid-term (second) exam? Is the second mid-term exam (public exam) better/ worse than the first mid-term exam? Please explain.''}
\end{center}
\begin{center}
\textit{
``What suggestions do you have to improve the Public Exam in the future?''}
\end{center}

\noindent
Examples of student responses included:
\begin{center}
\textit{
``I think it is great and it helps us student prepare what may be on the exam.'' [sic]}
\\ \, \\
\textit{
``I think there is a lot of benefit. I feel more prepared and less anxious.''}
\\ \, \\
\textit{
``It does benefit knowing some of the questions yet I still struggled.''}
\\ \, \\
\textit{
``I feel worse, I don't understand the point of having hidden content on public exam.''}
\\ \, \\
\textit{
``Maybe hide answers but show values during the public. Then change the values and figures during the exam.''}
\\ \, \\
\textit{
``Provide the equations for each questions.''}
\\ \, \\
\textit{
``Don't withhold content of or parts of the questions - this makes it very hard to know how and what to solve. Only withhold values and numbers. ''}
\end{center}

\noindent
The percentage of public questions on each exam varied over time, in response to student feedback.
On the first midterm, 50\% of the exam questions were made public ahead of time, and that fraction increased to 57\% and 60\% on the second midterm and final exam, respectively.  One open-ended question asked about this directly:%
\begin{center}
    \textit{
    ``Does increasing the percentage of Public Exam questions compare to the last two midterms help you?''}
\end{center}
Student responses to this question were mostly positive, with typical responses being:
    \begin{center}
    \textit{
    ``Yes! I liked being prepared to the best of my ability and knowing what to study helped a lot.''}
    \\ \, \\
    \textit{
    ``Yes, I feel more confident going into midterm 2 and the final than I did going into midterm 1.''}
    \\ \, \\
    \textit{
    ``It did help but [I] still struggled.''}
    \\ \, \\
    \textit{
    ``I think it did help, it reduced my anxiety.''}
\end{center}

We first coded comments related to Public Exams as either positive or negative.  Fig.~\ref{f:like/dislikes} shows the fraction of students whose responses included a positive comment about Public Exams and the fraction whose responses included a negative comment about Public Exams.  (A student who includes both positive and negative comments would be counted in both tallies.)  Initially, 92.31\% of students had a positive comment about public exams, while there were no negative comments about public exams at all.  After the second midterm, there was a significant drop in the fraction of students with positive comments about public exams to 58.82\%, as well as the first emergence of negative comments from 8.82\% of students.  At the time of the final exam, the fraction of students giving a positive comment was the highest (97.44\%), but the fraction of students giving a negative comment also increased (15.38\%).

Overall, students reported they liked the exam format and identified positive impacts such as anxiety reduction.  For all three exams, the fraction of students giving positive comments about public exams was much larger than the fraction of students giving negative comments about public exams.  The initial unanimously positive reaction by students after the first midterm indicates a positive first impression about the public exam system.  After the second midterm, there is a systematic shift toward negative comments: both a decrease in the fraction of students giving a positive comment and an increase in students giving a negative comment.  This may indicate the wearing off of the initial optimism as students encounter problems with the exam system; it may also be related to an increase in difficulty of the subject matter.  Interestingly, both the fraction of students giving positive comments and the fraction of students giving negative comments were highest at the time of the final exam. The significant increase in the fraction of students giving a positive comment may reflect their adjustment to the public exam system and growing confidence in how to use it well. Meanwhile the increase in the fraction of students giving a negative comment also suggests that a minority of students encountered persistent challenges with the public exam system. Students also consistently requested a higher percentage of public questions on the exams. Thus, while the overall attitude toward public exams is positive, we see a meaningful trend over time that shows student progression from initial optimism, through a period of adaptation. After gaining experience with it, students' perception of the public exam system remained highly positive, but with some substantive criticism.

As illustrated in the quoted student responses above, the Public Exam system acts like a study guide and benefits students by helping them focus their preparation efforts on the relevant material (directing to core concepts).  Even with some information withheld, students knew at least the content on which they would be assessed. Students also noted benefits to confidence and reduced anxiety.  However, they also expressed dissatisfaction about the withheld information on Public Exam questions and were unable to anticipate what would be asked in the real exam. Some students explicitly suggested to provide complete information about the problem in the Public Exams, but with a change in numerical values in the actual exam. They also suggested providing the equation that maps to each problem.

\begin{figure}[t]
    \centering
    \includegraphics[width=0.65\linewidth]{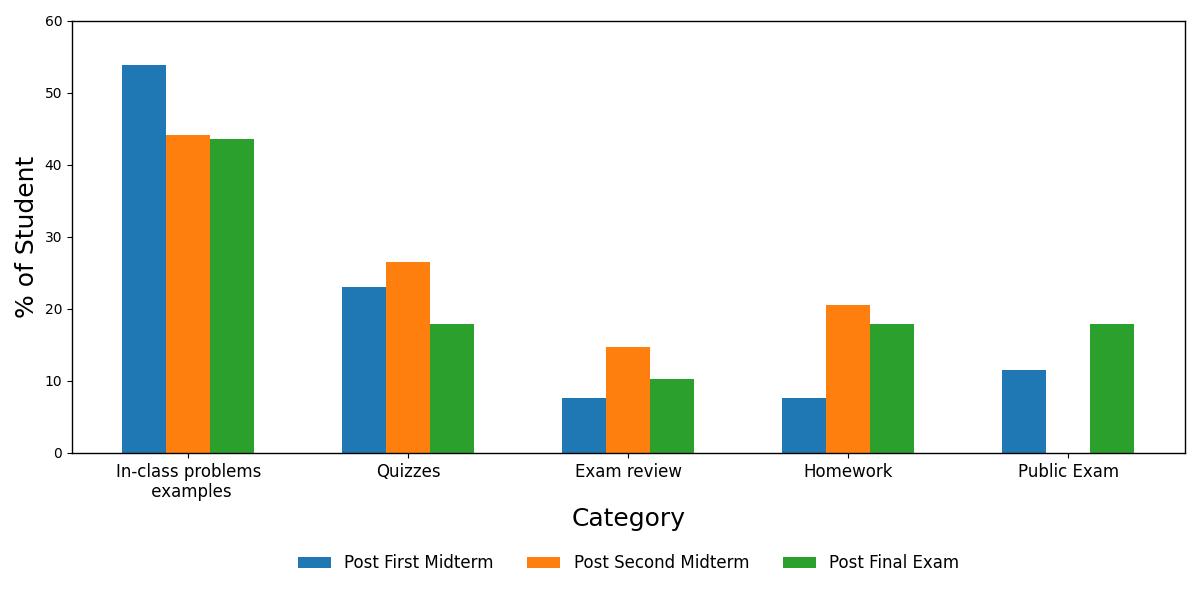}
    \caption{Fraction of students who mentioned a given item in open-ended responses to the question ``What has been most helpful for your learning in this class so far? (e.g., Describe the time(s) in this class when you were most engaged.)''}
    \label{f:helpful}
\end{figure}

Aside from analyzing positive and negative comments, we also specifically looked for student comments that addressed the themes previously identified by Wiggins et al. - language issues, directing to core concepts, deepening thought, authentic engagement, anxiety or confidence, and collaboration \cite{wiggins2022public}.  We found a significant number of student comments relating to directing to core concepts and to reducing anxiety; a detailed statistical analysis of these responses is conducted in Sec.~\ref{sec:Stats} below.  We did not find any student responses that directly addressed the impact of public exams on language issues or on collaboration.  For deepening thought and authentic engagement, there was one relevant student comment but insufficient data to perform a statistical analysis.  The student comment relating to deepening thought was
\begin{center}
\textit{
``I would prefer having more public questions because I learn these concepts much more deeply and feel more prepared to take the exam.''}
\end{center}
Similarly, the one comment addressing authentic engagement was 
\begin{center}
\textit{
``I was most engaged when we were going over practice problems and the public exam.''}.
\end{center}
%

%
\subsubsection{Comments Regarding Classroom Practices}
%

Students were also asked about classroom practices in various ways.  One question focused on student engagement:
\begin{center}
    \textit{
    ``What has been most helpful for your learning in this class so far? \\ (e.g., Describe the time(s) in this class when you were most engaged.)''}
\end{center}
Examples of student responses are as follows, and the most common responses are tabulated in Fig.~\ref{f:helpful}:
\begin{center}
    \textit{
    ``I was most engaged when we were going over practice problems and the public exam.''}
    \\ \, \\
    \textit{
    ``Taking quizzes and when the prof. uses simulations to explain concepts.''}
    \\ \, \\
    \textit{
    ``The homework has helped me lot and using a tutor to simplify physics.''}
\end{center}
All five of the top responses to this question referred to practical applications of procedural knowledge: in-class examples and practice problems, quizzes, homework problems, exam review, and Public Exam questions.  Other, less common responses included opportunities to work together with classmates, interaction with the instructor, visualizations and animations, and the quality of lecture notes.  Note that we separately tallied mentions of Public Exams from mentions of in-class problems and examples, but there could potentially still be some overlap between the two categories. It would be interesting to separate the two explicitly in a future study.

\begin{figure}[t]
    \centering
    \includegraphics[width=0.80\textwidth]{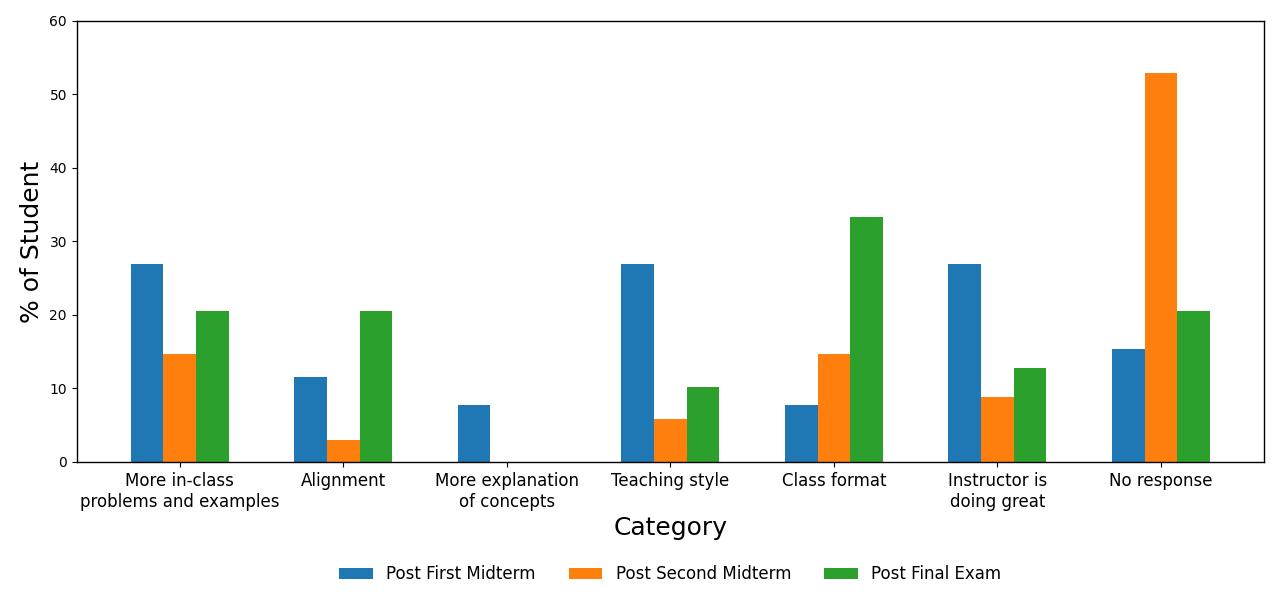}
    \caption{Open ended responses for the suggestions to the instructor to make this course better.}
    \label{f:Suggestions}
\end{figure}

The heavy emphasis on applying procedural knowledge in student responses to this question may reflect the discipline-specific emphasis in physics.  Problem solving by applying a specific mathematical procedure is a hallmark of physics, and the students' prioritization of problem-solving skills shows that they understand the key cognitive processes on which they will be assessed.  This further suggests that the public exam system could be adapted to a physics context by heavily emphasizing the procedural aspects of a problem.  As noted by a student response above, it is possible to use a \textit{fully revealed} public question and just change the numbers to assess the application of procedural knowledge.

Another question entailed an open-ended request for suggestions to make the course better:
\begin{center}
\textit{
``What are your suggestions to Instructor in order to make this course better? \\ This can be about anything, such as the organization of the course or the way of teaching.''}. 
\end{center}
Examples of student responses are as follows, and the most common responses are tabulated in Fig.~\ref{f:Suggestions}:
\begin{center}
\textit{
``I think there should be more opportunity for practice problems.''}
\end{center}
\begin{center}
\textit{
``Keep doing what you're doing, you are amazing.''}
\end{center}
\begin{center}
\textit{
``More class activities, sometimes I had a difficult time paying attention \\ and sitting through an entire lecture can be pretty hard.''}
\end{center}
Many students did not offer meaningful suggestions, with 15.38\%, 52.94\%, and 20.51\% of the students giving no response to this question at all following the first midterm, second midterm, and final exam, respectively. Similarly,  26.92\%, 8.82\%, and 12.82\% of students indicated that the Instructor is doing great and had no further suggestions. Among students who did offer suggestions, the most common suggestions included giving more in-class examples and practice problems and increasing the alignment of lectures and assessments, as well as comments related to the instructor's teaching style and class format.  Again, the request for more concrete example problems reflects an emphasis on applying procedural knowledge characteristic of Physics as a discipline.

Finally, on the surveys given after the two midterm exams, students were also asked a closely-related question about their suggestions to enhance the learning experience:
\begin{center}
\textit{
``What suggestion(s) can you make that would enhance your learning experience in this class?''} 
\end{center}
\begin{figure}[t]
    \centering    
    \includegraphics[width=0.80\textwidth]{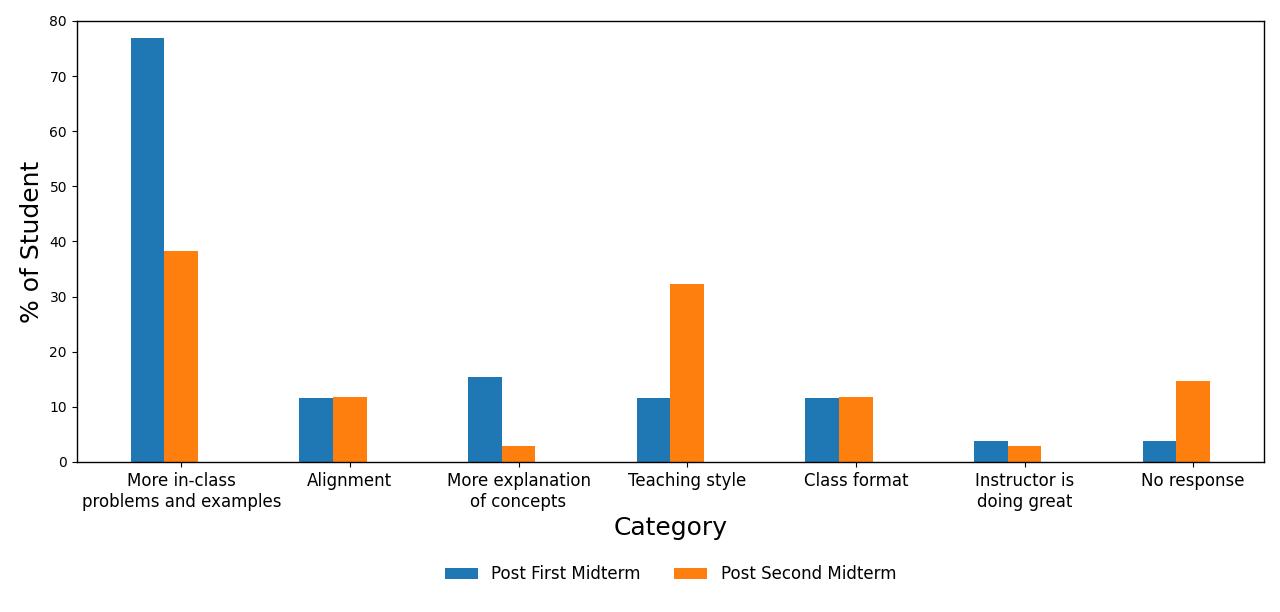}
    \caption{Open ended responses for the suggestions of the students that can enhance their learning experience in the class.}
    \label{f:Enhance}
\end{figure}
Examples of student responses are as follows, with the most common responses shown in Fig.~\ref{f:Enhance}: %
\begin{center}
    \textit{
    ``I thought that by having more example questions benefitted me when learning the concepts.''}
    \\ \, \\
    \textit{
    ``Do more in class examples and the questions that are on the lecture to provide the answers so we can look at the answers.''}
    \\ \, \\
    \textit{
    ``More homework or practical examples with steps.''}
\end{center}

Again, the most common suggestion was to increase the number or depth of in-class problems and examples (76.92\% of the students following the first midterm exam, decreasing to 38.24\% of students following the second midterm). The other most common responses included more explanation of concepts and increasing the alignment of the homework with the lectures. Additional suggestions included more interactive class activities, more student feedback, and practice exams which only change the values compared to the actual exams.  The range of responses shown in Figs.~\ref{f:Suggestions} and \ref{f:Enhance} highlight the diverse learning needs of NMSU students and the importance of applying procedural knowledge in physics.

%
\subsection{Results:  Statistical Analysis of Qualitative Data}
\label{sec:Stats}

We next performed a statistical analysis of the frequency of positive versus negative comments similar to the procedure described by Wiggins et al. \cite{wiggins2022public}.\footnote{
We note that ``positive'' and ``negative'' may refer either to a student's affective response (opinion) about the Public Exams or to the impact on a student's learning outcomes.  The standard example mentioned by Wiggins et al. \cite{wiggins2022public} is the possibility of a comment like ``I hate public exams because they force me to think more deeply,'' which would indicate a positive impact on learning but a negative affective reaction.  In our study such conflicting cases did not occur; all students who expressed a positive affective response to the public exams connected it to a positive impact on their learning, and all students who expressed a negative affective response connected it to a negative impact on their learning.  Thus, while the intention of coding the open-ended responses was to identify positive or negative impacts on student learning, in our study there was no difference between this and coding for positive or negative affective response.
}
Based on the student feedback provided in the open-ended responses, we counted the number of positive and negative comments on a student by student basis.  The primary metric was the ratio of positive comments to total comments in a given category.  Comments related to the Public Exam system were used as the test sample, while comments about topics unrelated to Public Exams served as the control group. We then conducted a Pearson's $\chi^2$ test of independence to assess whether students' positive-comment rate on Public Exams was independent of their positive-comment rate on the control sample.  Thereafter we conducted a two-sample, two-sided T test to assess whether the difference in positive-comment rate between Public Exams and the control sample was statistically significant.

We emphasize that in our procedure, we divide the number of positive comments by the \textit{total number} of comments (positive plus negative).  This is a small but significant difference in methodology compared to Wiggins et al. \cite{wiggins2022public}, in which ``the results are presented as [the ratio] ‘PositiveInstances:NegativeInstances’.''  This matters for the Pearson independence test, which needs a valid ratio for both the Public Exams category and the control category, on a student-by-student basis.  A student who comments only on the Public Exams, but does not have any comments in the control group, does not have a valid pair of ratios that can be used in the correlation analysis.  In our case, we have 39 students in the dataset, and only one student was dropped from the correlation analysis due to having zero comments in the control group.  Therefore, using a ratio of positive comments to total comments only loses 2.56\% of students from the correlation test in our dataset.

But since Wiggins et al. \cite{wiggins2022public} uses a ratio of positive comments divided by \textit{negative} comments, that ratio becomes undefined much more frequently (whenever a student gives only positive comments).  Conversely, only a student who offers both a negative comment related to Public Exams and a negative comment included in the control group would be counted in a Pearson test for independence.  If we had used the positive to negative ratio as in Wiggins et al. for our analysis, we would have lost 95\% of the total dataset, which would have hugely biased the resulting statistical test.

We proceed in the spirit of Wiggins et al.'s statistical analysis, but using the positivity rate (positive divided by total comments) rather than the positive-to-negative ratio.  We also revisit Wiggins et al.'s published data using our statistical method in Sec.~\ref{sec:Wiggins} to allow for a fair comparison of the results.

%
\subsubsection{Fraction of Positive Comments: Public Exams vs. Control}
%

Representative comments about students' positive and negative opinions about Public Exams were given in Sec.~\ref{sec:PEcomments} above.

\begin{figure}[p]
    \centering
    \includegraphics[width=\textwidth]{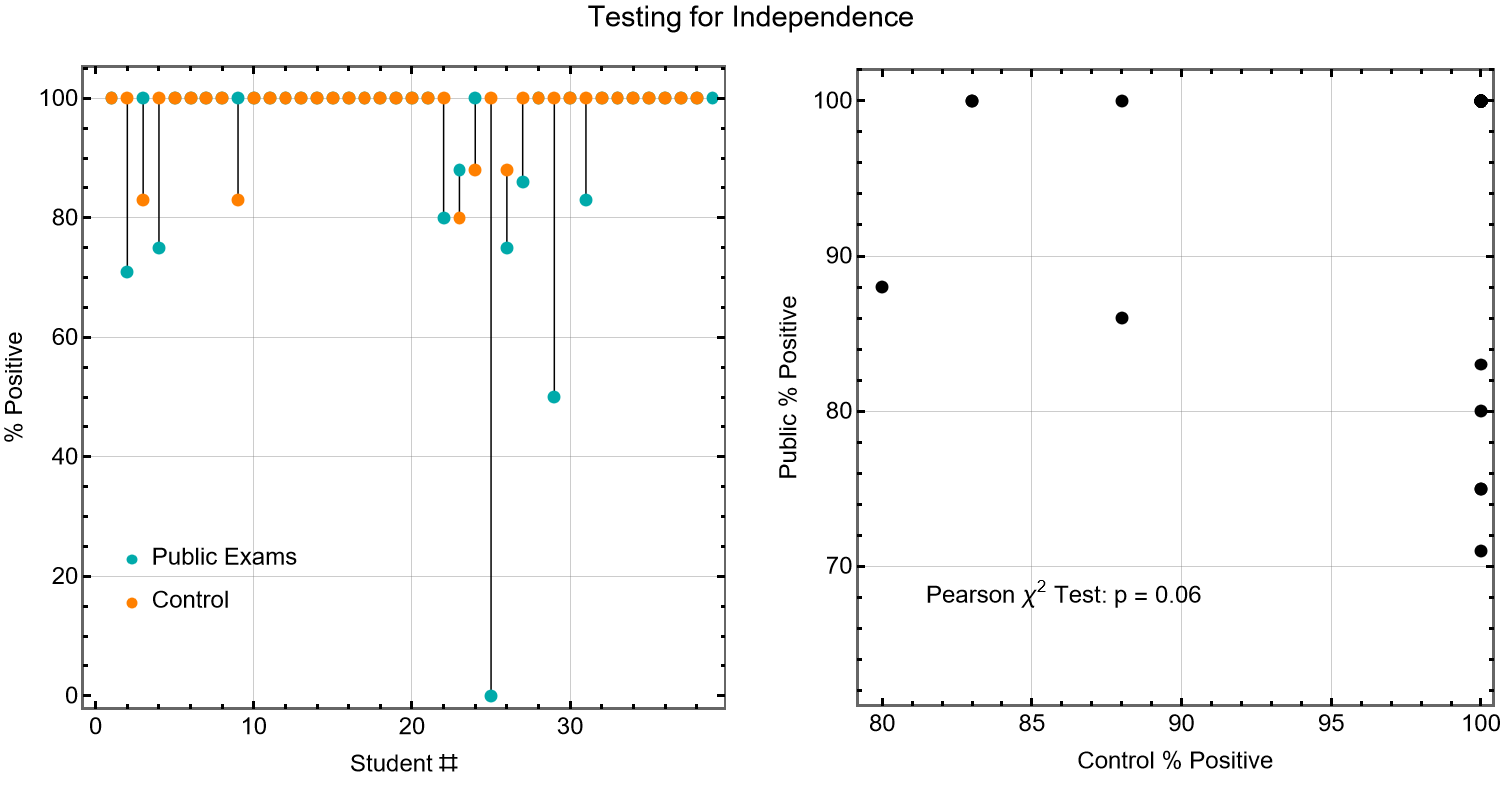}
    \caption{Scatter plot for positive responses of students for Public Exams vs control group.}
    \label{f:correlation_test}

    \vspace{2cm}

    \includegraphics[width=0.7\textwidth]{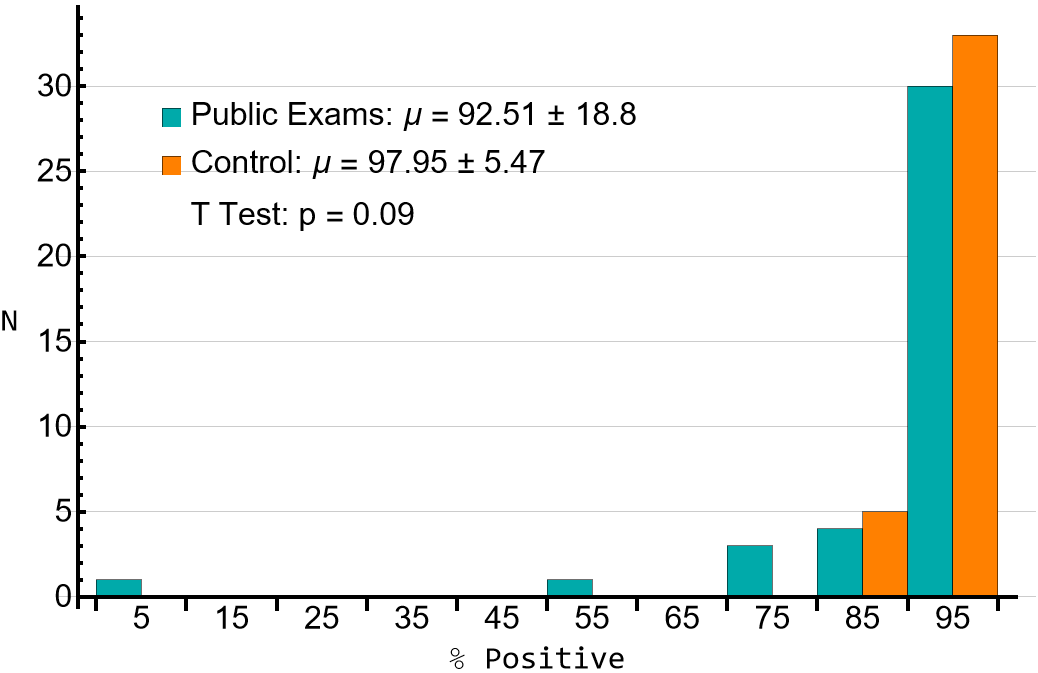}
    \caption{T-test for the positive comments of Public Exam vs Control.}
    \label{f:T_test}
\end{figure}

In Fig.~\ref{f:correlation_test}, we show the correlated student positivity rates for the Public Exam and Control categories.  A Pearson's $\chi^2$ test for independence resulted in a $p$-value of 0.06.  While this does not reach the 5\% threshold necessary to reject the null hypothesis with high confidence, it does suggest that the positivity rate for Public Exams is independent from the positivity rate for the control group at the 10\% level.  That is, students' expressed opinions are generally different for the Public exam from the control group; they are not just repeating or copying the same behavior. 

We then conducted a T-test, which includes even students who were dropped from the Pearson test, to compare the difference in means and its significance.  The histograms of the Public Exam and Control positivity rates are shown in Fig.~\ref{f:T_test}. 
From the T-test, the mean positivity rate for the Public Exam category ($\mu = 92.51, \sigma = 18.8$) was lower than for the control group ($\mu = 97.95, \sigma = 5.47$), with a $p$-value of 0.09.  This evidence suggests that there may be a meaningful difference between the positivity rate for Public Exams compared to the control group, and surprisingly this data suggests that the Public Exams could actually be \textit{less positive} than the control.  We can clearly see from Fig.~\ref{f:T_test} that, while a high positivity rate is quite common for Public Exams, there are a significant number of students with overall negative responses.  
This is in contrast to the significantly \textit{higher} positivity rate seen in Wiggins et al.: a positive to negative ratio of $97:22 = 4.41$ for Public Exams versus $74:38 = 1.95$ for control.  In our data, the positivity rate for the Public Exams may actually be lower than for control (although still highly positive overall).

%
\subsubsection{Directing to core concepts: Public Exams vs. Control}
%

We found a significant number of student comments (39) that referred to the impact of Public Exams on helping to direct them to core concepts, both positively and negatively.  Representative comments include:
\begin{center}
\textit{
``I like that we were able to see the format and the wording of the questions. \\ This causes problems in traditional exams, I like knowing what to expect and prepare properly.''} 
\end{center}
\begin{center}
\textit{
``I honestly think it is good. It was a nice ``study guide'' to help prepare us on the actual exam. \\ If we were to do in a traditional exam then we wouldn't have an idea of what is going to be on it.''} 
\end{center}

\begin{figure}[p]
    \centering
     \includegraphics[width=\textwidth]{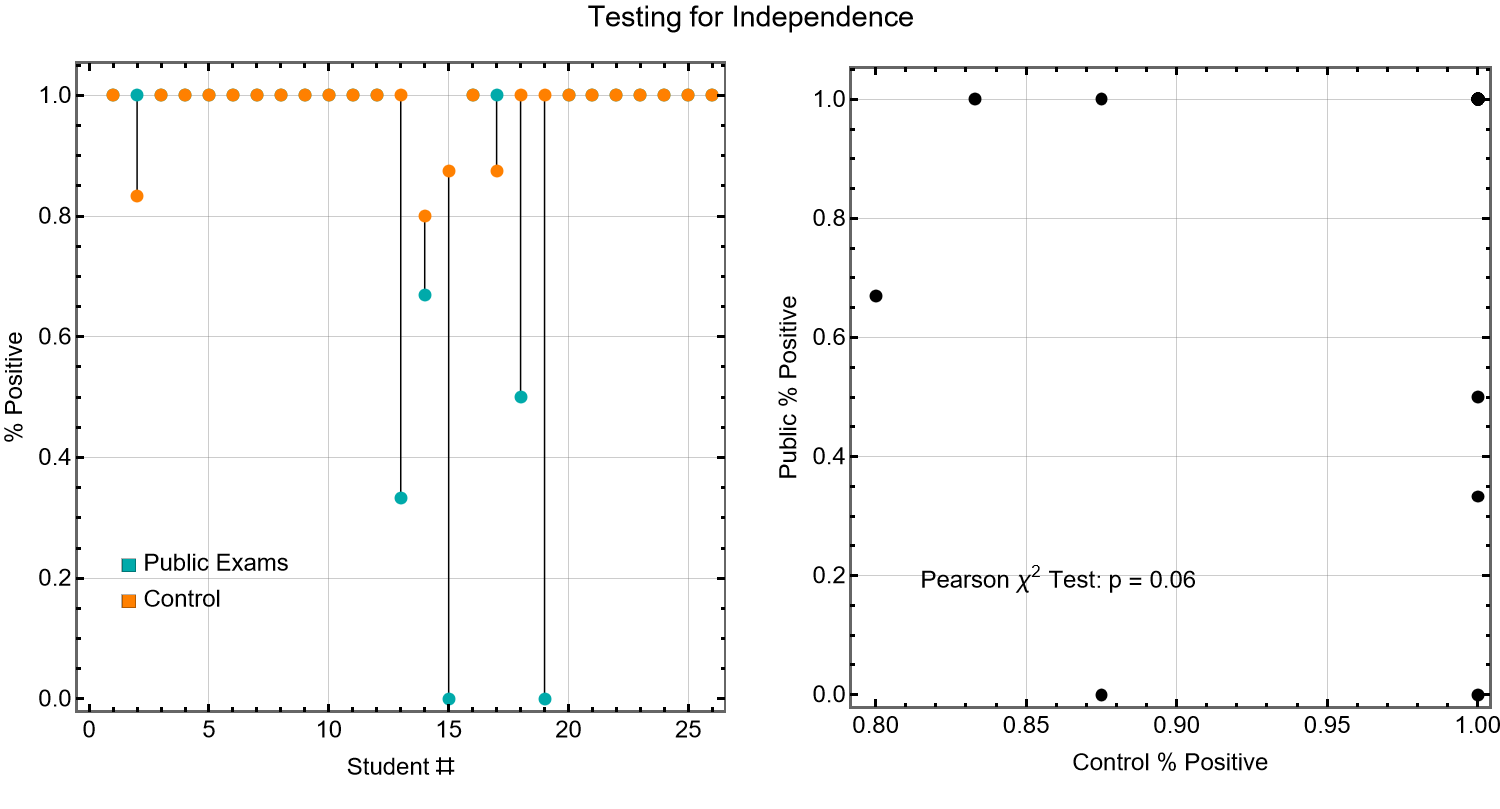}
    \caption{Scatter plot Public Exam vs control for Directing to Core Concepts.}
    \label{f:Core_Concepts_P_Test}

    \vspace{2cm}

    \includegraphics[width=0.7\textwidth]{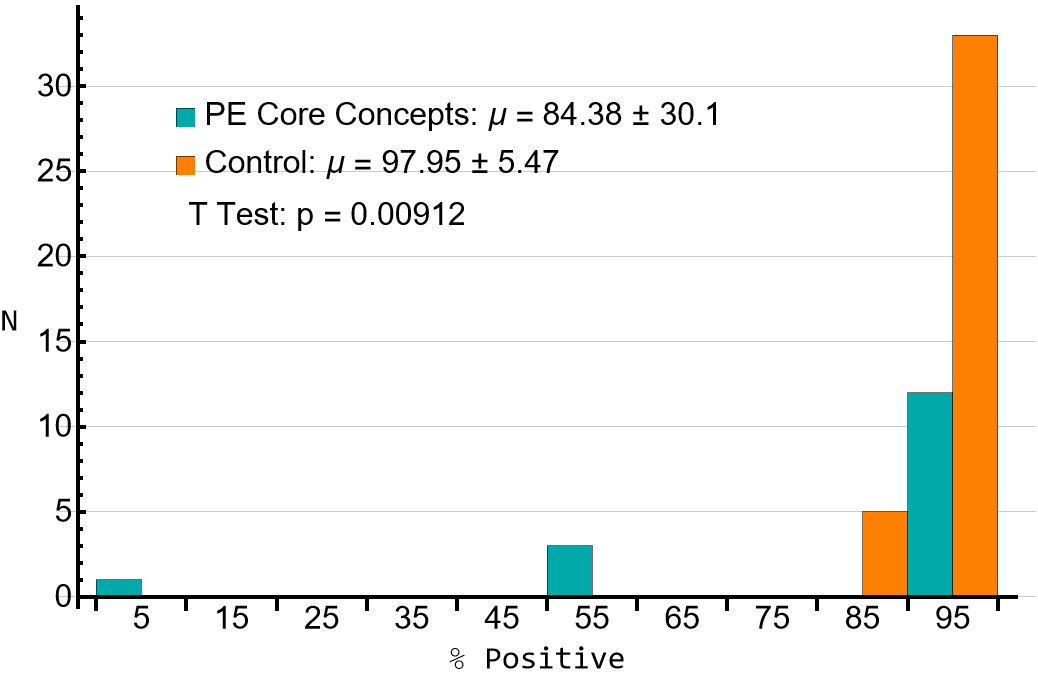}
    \caption{T-test for Public Exam vs control for Directing to Core Concepts.}
    \label{f:Core_Concepts_T_Test}
\end{figure}

As with the overall positivity rates for Public Exams and the control group, we repeated the $\chi^2$ test (Fig.~\ref{f:Core_Concepts_P_Test}) and T test (Fig.~\ref{f:Core_Concepts_T_Test}) for these comments related to ``directing to core concepts.''  For this data, the $\chi^2$ test returned $p = 0.06$, nearly identical to the value for the overall opinion comparison.  Again, this shows that students' opinions regarding Public Exams were reliably independent of their opinions on the control group.

The mean positivity rate for Public Exams helping direct students to core concepts $(84.4 \pm 30.1)\%$ was again lower than the positivity rate for control $(98.0 \pm 5.5)\%$.  The change in positivity rate for Public Exams directing to core concepts compared to control was highly statistically significant: $p = 0.009$.  This again contrasts with the overwhelmingly positive response to ``directing to core concepts'' in Wiggins et al., where the positive-to-negative ratios were $55:5 = 11.0$ for Public Exams and $74:38 = 1.95$ for control.  Again, while students still had a high positivity rate for directing to core concepts ($84.4\%$), the change relative to the control group is significantly reduced compared to Wiggins et al.

%
\subsubsection{Reducing anxiety: Public Exams vs. Control}
\label{sec:anxiety}
%

In our study, we had a total of 9 comments relating to the impact of Public Exams on test anxiety.\footnote{We note that anxiety is a complicated response that is the subject of substantial research in its own right\cite{alenina2026dual, vonderembse2018test}.  Anxiety functions as a complex dual-structure hierarchical system rather than treating it as a unitary construct\cite{alenina2026dual}. General anxiety ranges on a continuous scale from everyday milder traits to chronic clinical disorders, while domain-specific anxieties emerge in specific settings such as math anxiety, test anxiety etc.  In the open-ended responses discussed here, we do not attempt to specify what the nature of students' anxiety experience is; we simply rely on students' self-reported experiences.  In this sense, the meaning of ``anxiety'' depends on the individual student's perception.  This is in contrast with measures of student anxiety obtained from the MSLQ surveys, discussed later.}
In response to the question
\begin{center}
\textit{
``What is your opinion about the public exam? Does it potentially benefit from the traditional exam? \\ Do you have any suggestions for any kinds of exams in general?''}
\end{center}
six students commented positively, with a typical comment being 
\begin{center}
\textit{
    ``I think it did help, it reduced my anxiety''} \: ,
\end{center}
while one student commented negatively 
 \begin{center}
\textit{
    ``I personally hate public exams but I hate public exams more \\ those with test anxiety, do not do well with public exams''} [sic].
\end{center}
This same student reported a consistently negative opinion of public exams.  In response to the question 
\begin{center}
\textit{
``Does increasing the percentage of public exam questions compare to the last two midterms help you?''}
\end{center}
the student's answer was
 \begin{center}
\textit{
``It was way more difficult with more questions. We will see how the score come out''}[sic].
\end{center}
When given an open-ended chance to suggest changes,
\begin{center}
\textit{
``What suggestions do you have to improve the public exam in the future?''}.
\end{center}
this student's response,
 \begin{center}
\textit{
``Open note, or able to do exam with others seeing we collab on quizzes''} [sic]
\end{center}
suggested alternatives to Public Exams altogether.  This student's pattern of responses suggest that the public exam system actually \textit{increased} their test anxiety.

\begin{figure}[p]
    \centering
    \includegraphics[width=\textwidth]{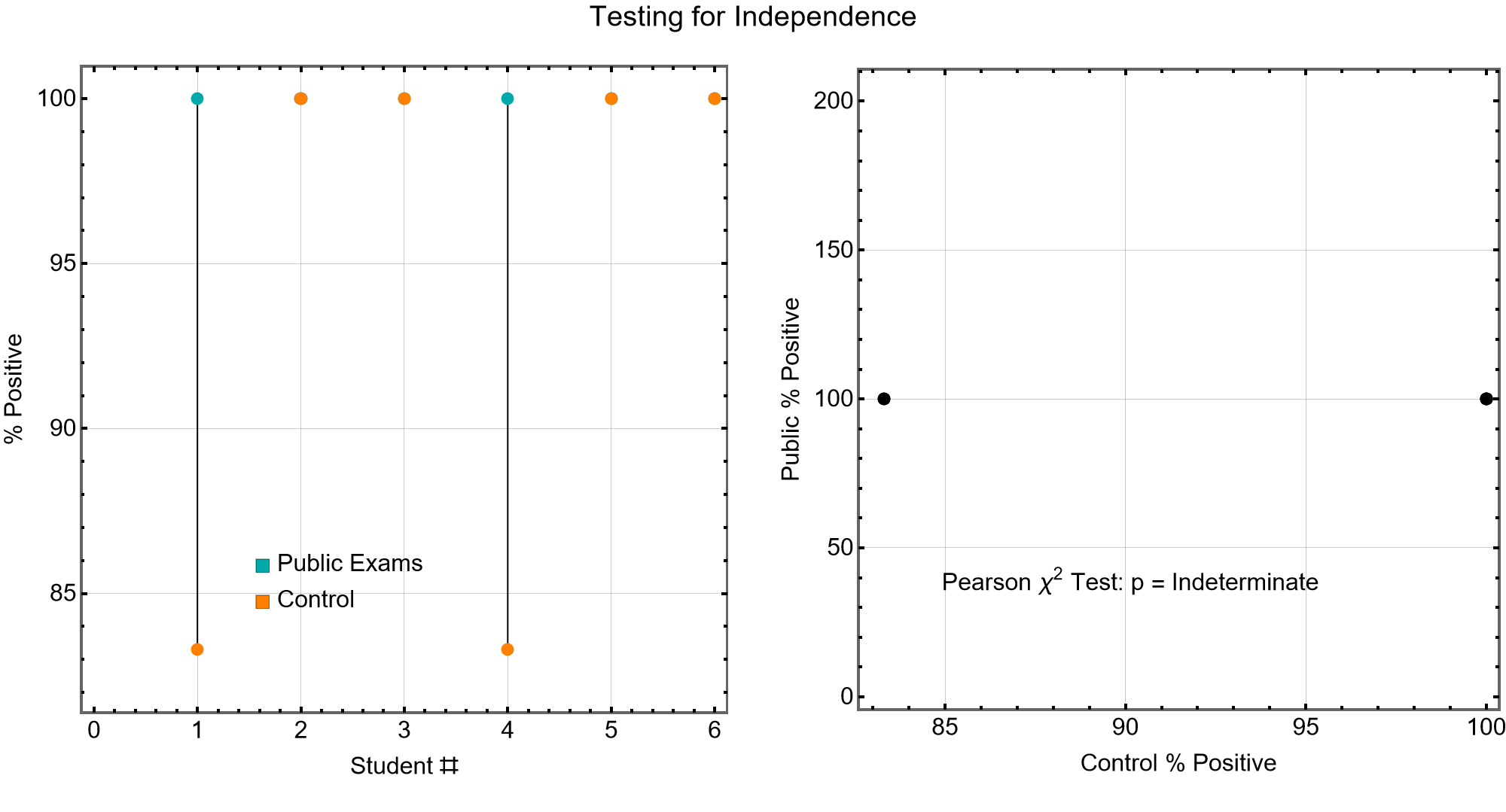}
    \caption{Scatter plot Public Exam vs control for Reducing Anxiety.}
    \label{f:Anxiety_P_Test}

    \vspace{2cm}

    \includegraphics[width=0.7\textwidth]{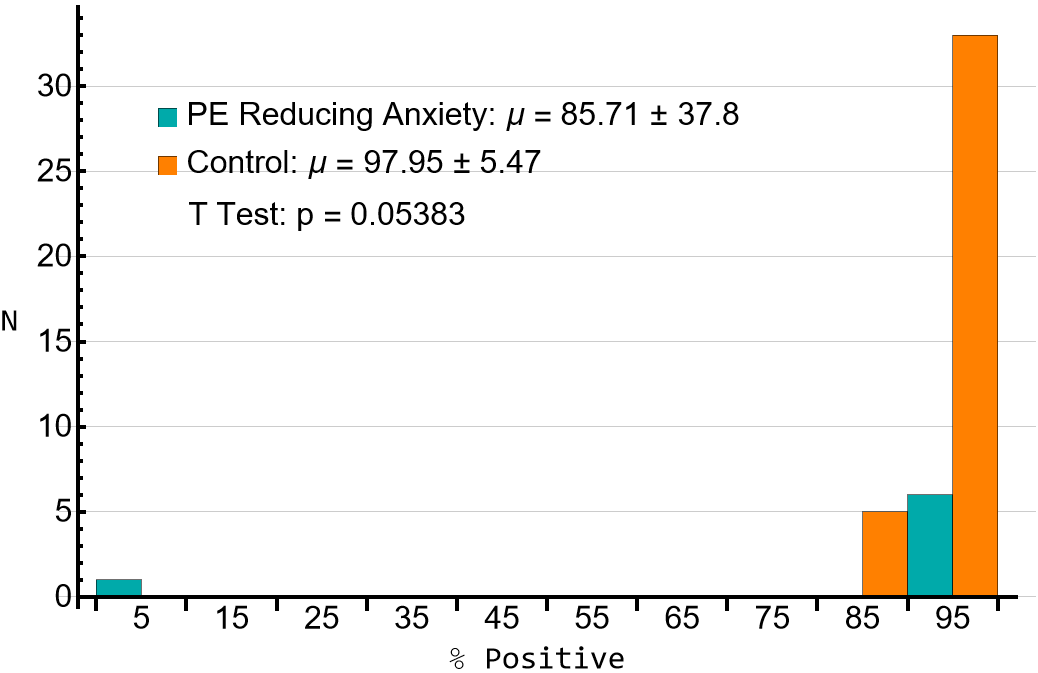}
    \caption{T-test for Public Exam vs control for Reducing Anxiety.}
    \label{f:Anxiety_T_Test}
\end{figure}

A Pearson $\chi^2$ test for independence on this small sample was inconclusive, but the mean positivity rate for comments about Public Exams' impact on anxiety was $(85.7 \pm 37.8)\%$ while the positivity rate for control was $(98.0 \pm 5.5)\%$ shown in Fig.~\ref{f:Anxiety_P_Test}. The change in positive comments about Public Exams' impact on anxiety compared to control was statistically significant ($p = 0.05$) shown in Fig.~\ref{f:Anxiety_T_Test}. Again, this stands in contrast with the results of Wiggins et al., where the positive-to-negative ratio for Public Exams was $31:4 = 7.75$, while for control it was $74:38 = 1.95$. Consistently, students in our analysis reported overall positive responses to the Public Exams across these various categories, with a positivity rate that is significantly lower than seen in Wiggins et al., and lower even than the control group.

%
\subsection{Re-Analysis of Wiggins et al.}
\label{sec:Wiggins}
%

\begin{figure}[p]
    \centering
    \includegraphics[width=\textwidth]{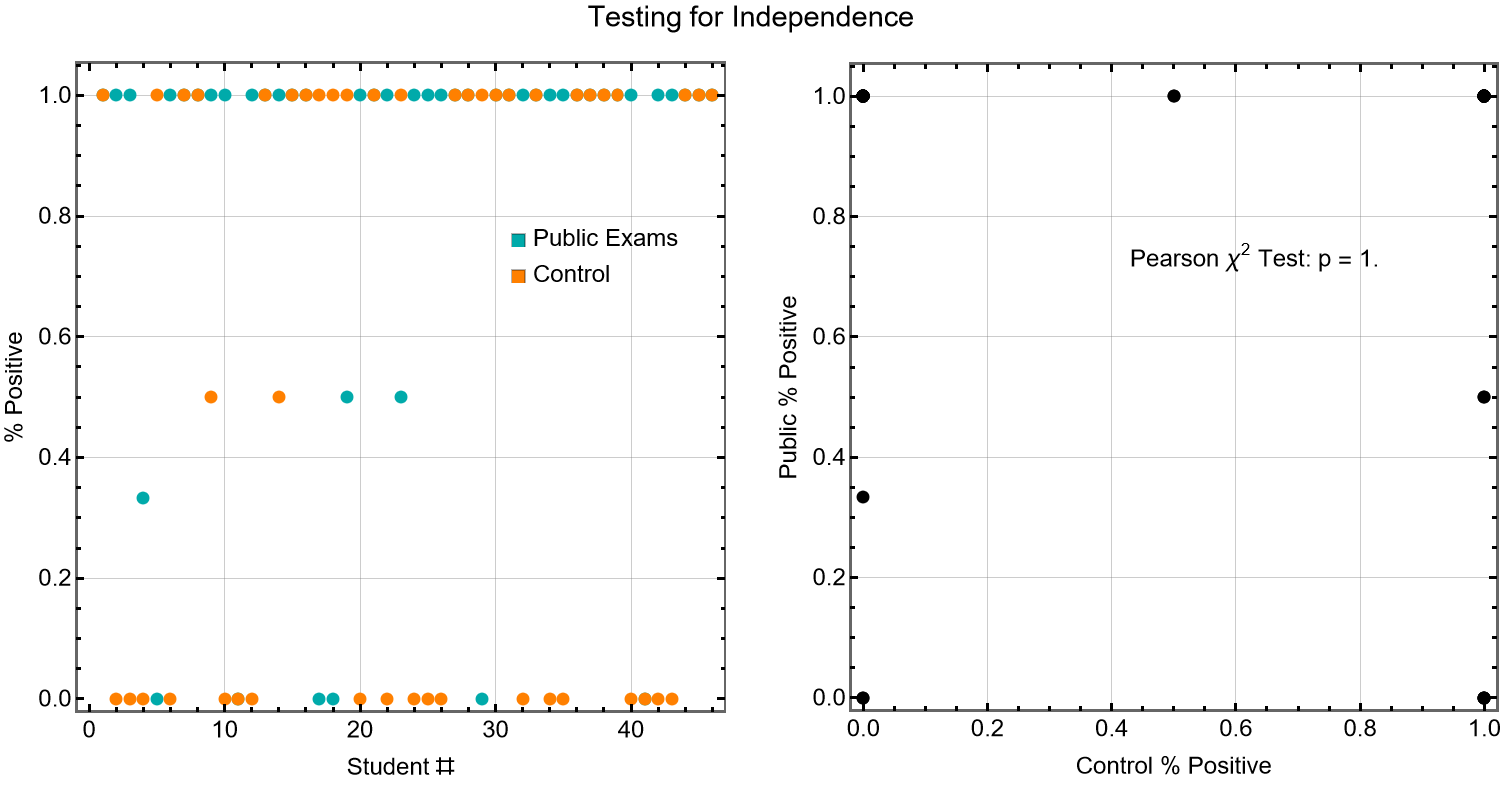}
    \caption{Scatter plot Public Exam vs control for Wiggins data.}
    \label{f:scatterplot_wiggins data}

    \vspace{2cm}

    \includegraphics[width=0.7\textwidth]{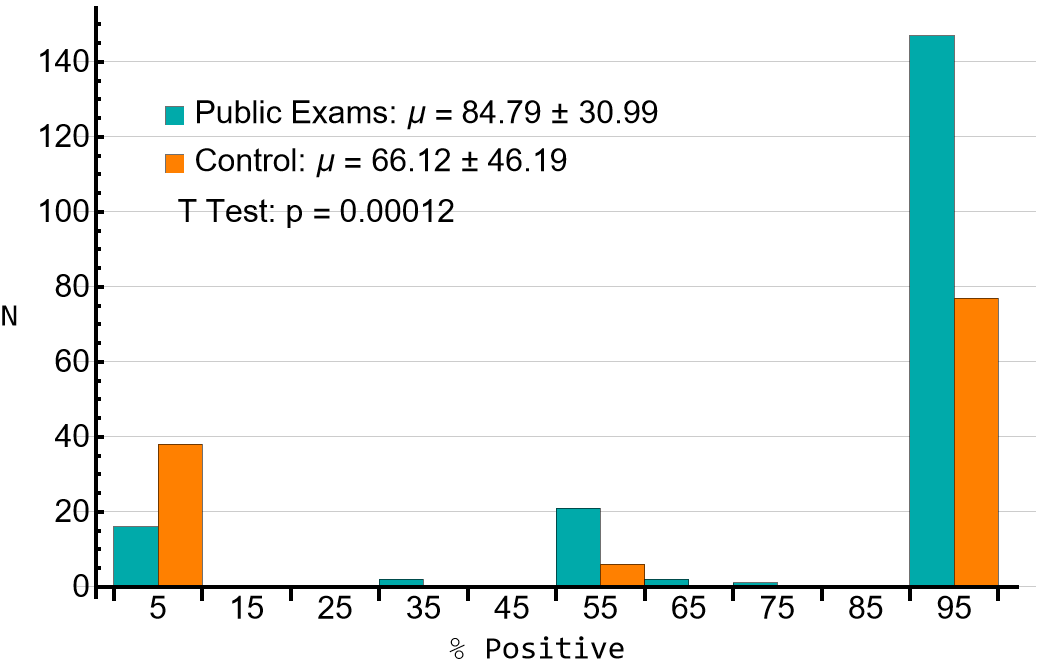}
    \caption{T-test for Public Exam vs control for Wiggins data.}
    \label{f:histogram_wiggins data}
\end{figure}

To perform an apples-to-apples comparison with the results of Wiggins et al. \cite{wiggins2022public}, we reanalyzed the publicly available source data accompanying Ref.~\cite{wiggins2022public} in the same way as for our own data above.  If we attempted to use a student-by-student test for independence using the positive-to-negative ratio, only $1.1\%$ of students would have remained in the data set, leading to an inconclusive result.  Instead, we used the positivity ratio as in the analysis of our own data.   

The $\chi^2$ test for independence (Fig.~\ref{f:scatterplot_wiggins data}) returned $p = 1.0$, demonstrating a remarkable amount of correlation among student responses (often either $100\%$ positive or $100\%$ negative).  The mean positivity rates for Public Exams and control were $(84.8 \pm 31.0)\%$ and   $(66.1 \pm 46.2)\%$, respectively, and the T-test analysis (Fig.~\ref{f:histogram_wiggins data}) returned $p = 0.00012$, indicating that this difference is highly significant.  

Thus our methodology, applied directly to the publicly available data of Ref.~\cite{wiggins2022public}, confirms their conclusions and provides a meaningful baseline for comparison to our own data.  We do see that the mean positivity fraction for comments regarding Public Exams in Wiggins et al.'s study was significantly higher than for control, at a very high level of significance.  This highly positive result from Wiggins et al. in a biology context further underscores the differences we observe in a physics context.

%
\subsection{Results:  Other Quantitative Data and Mixed-Methods  Analysis}
%

\begin{figure}[t]
    \centering
    \includegraphics[width=1\textwidth]{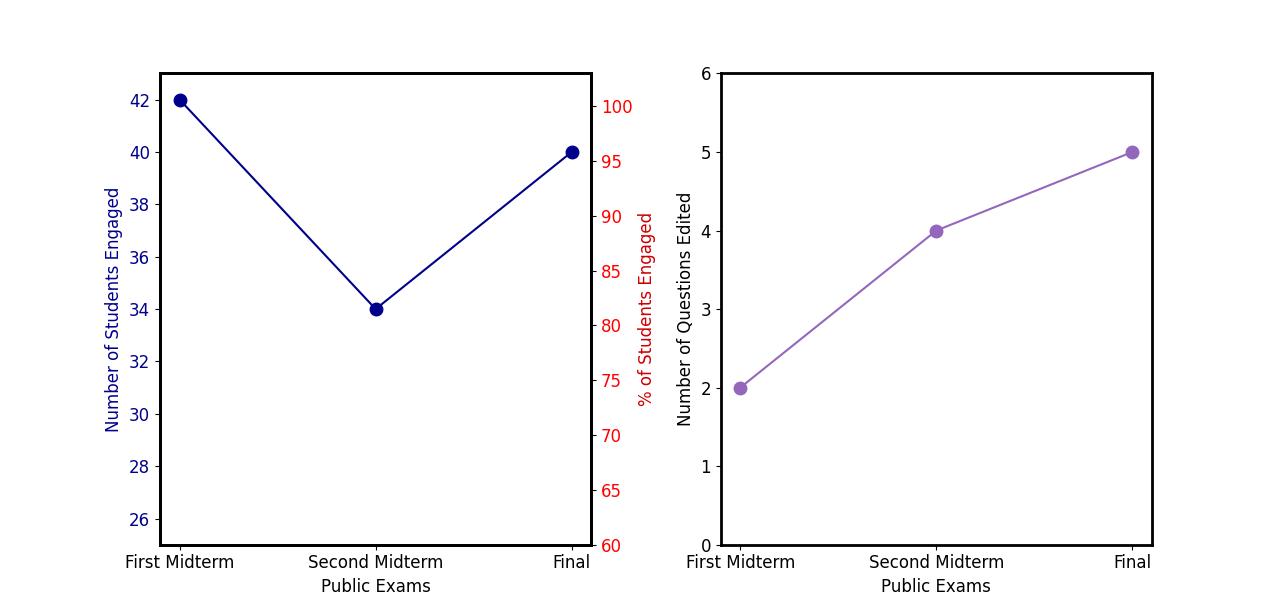} 
    \caption{Left: Number and percentage of students who engaged with the Public Exam process for each exam.  Right: Number of questions edited for each exam.  Data is only available for the 42 individuals who consented to participate in this study.}
    \label{f:Engagement}
\end{figure}

One of the major findings of Wiggins et al. \cite{wiggins2022public} was a very strong correlation between the engagement rate of students in the process of submitting edits to the public exams and their overall course grades.  This evidence supported their hypothesis that Public Exams promote ``authentic engagement.''  As described above, students in the class we study were offered a small extra credit incentive for ``engaging'' with the Public Exam process.  The incentive consisted of 0.67\% extra credit to their overall course grade for each time they participated in the Public Exam editing process, for a maximum of 2\% extra credit if they participated all three times. 

To receive the extra credit incentive, students had to read the exam questions, and choose one of the following options:
\begin{enumerate}
    \item Edit a single question to improve that question, 
    \item Suggest a new/ different question or make a comment to generally improve the exam, or 
    \item Acknowledge that they have read the draft exam and have no comments. 
\end{enumerate}
Students received the full extra credit incentive no matter which of these three options they selected.  The data included in our study reflected only the amount of extra credit received on a given Public Exam, not which option they chose.  But although we do not know which option a given student chose, we can compare the versions of a given exam to see which problems changed during the revision process.

%
\subsubsection{Engagement Rate}
%

Fig.~\ref{f:Engagement} shows the counts of students who engaged in the Public Exam revision process for each exam, as measured by receiving the extra credit.  By this measure, the level of engagement with the Public Exams was extremely high across the three exams (100\%, 81\% and 95\%, respectively) and fairly consistent over time.  There is an interesting non-monotonic behavior in the engagement rate, however, with a decrease from the first midterm to the second midterm, followed by a significant increase on the final exam.  This pattern -- a high engagement rate, then a low engagement rate, and finally a medium engagement rate -- mirrors the pattern of students' net opinions about the Public Exams seen in Fig.~\ref{f:like/dislikes}.  Interestingly, these two measures are separated in time:  the students' engagement with the revision process as measured in Fig.~\ref{f:Engagement} occurred before the exam in question, while the opinions represented in Fig.~\ref{f:like/dislikes} were measured afterwards.  

In contrast, the number of edited questions increased monotonically over time, from 2 questions edited on the first midterm, to 4 on the second midterm, to 5 on the final exam.  Although the number of edited questions is small compared to the number of students in our study (11 total edited questions in a sample with 42 participants), the steady increase over time is noteworthy.

On the first midterm, students exhibited both a 100\% engagement rate and a unanimous positive opinion about Public Exams compared to traditional exams, but only 2 questions were edited.  This could be due to high student motivation at the beginning of the semester when confronted with a novel exam system intended to benefit them, and an opportunity to earn extra credit simply by participating.  But this initial enthusiasm did not translate into a high rate of editing questions, suggesting that the motivation had not translated into deep, ``authentic'' engagement yet.

On the second midterm, the engagement rate dropped to 81\% and the students' opinions became much more mixed (58.82\% of students made a positive comment about Public Exams, while 8.82\% of students made a negative comment), but the number of edited questions increased to 4.  The doubling in the number of edited questions suggests an increase in the rate of deep, ``authentic'' engagement with the Public Exams process, while the net decrease in student opinion and engagement rate (as measured by extra credit) suggests a cooling of student motivation for the Public Exams process.  Together, these may indicate that students were discovering some drawbacks and limitations of the Public Exam process as they became more familiar with it.

On the final exam, the engagement rate (measured by extra credit) increased back to 95\%, and students' opinions of the Public Exams improved significantly (97.44\% of students commenting positively and 15.38\% of students commenting negatively), while the number of edited questions again increased to 5.  In contrast to the competing signals on the midterm exams, here there was an increase in both editing problems and in students' opinions.  This recovery in opinion, along with a further increase in ``authentic'' engagement, suggests that students generally overcame the negative aspects they perceived about the Public Exams.  Although students retained their criticisms of the Public Exams (with negative comments increasing from 0\% to 8.82\% to 15.38\% over time), the recovery in positive opinion was even greater.  This pattern suggests that the students arrived at a compromise opinion, informed by their experiences about both the positive and negative aspects of Public Exams.

In future studies, it would be beneficial to change the options and/or incentive for suggesting edits during the Public Exam revision process.  For example, requiring students to submit a suggestion to receive the extra credit would make that grade a better measurement of ``authentic'' engagement.  Alternatively, the extra credit incentive could still apply to a ``no suggestions'' option, but future studies should track individually which option a student chose.

%
\subsubsection{Impact of Engagement on Exam Performance}
%

\begin{figure}[t]
    \centering
   \includegraphics[width=0.45\textwidth]{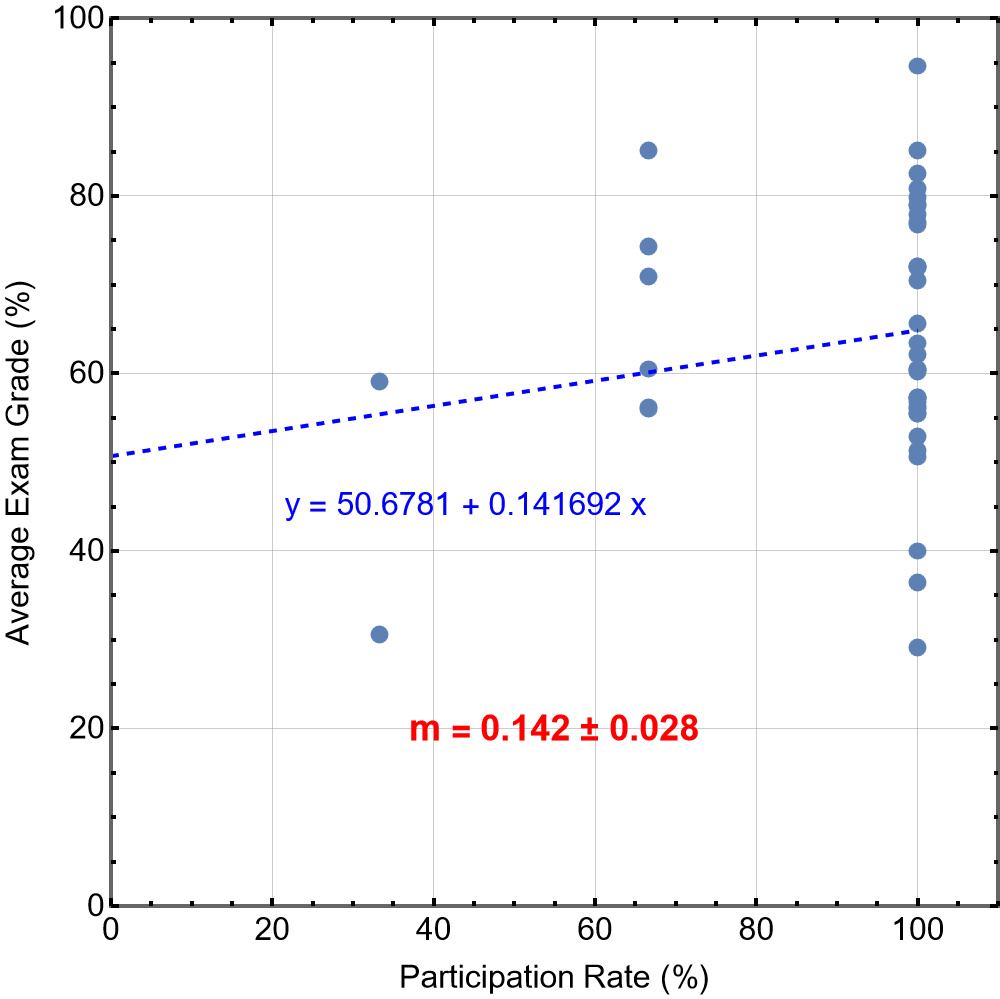}
    \caption{Scatter plot for number of times of participation on Public Exam versus average exam grade, with best-fit line and slope uncertainty shown.}
    \label{f:Performance}
\end{figure}

To study the impact of participation in the Public Exams revision process on student performance, we analyzed the correlation between a given student's engagement rate (as measured by receiving the extra credit) and their exam average (first midterm, second midterm and final exam, including extra credit).  This metric is somewhat different than in Wiggins et al. \cite{wiggins2022public}, which focused on the correlation of engagement with overall course grades (including homework and other assignments).  As noted above, the engagement rate data we have on an individualized basis reflects a ``shallow'' level of engagement -- a willingness to at least look at the draft exam and choose the ``no comments'' option.  For this reason in particular, it is far from clear whether we should expect to see the same level of strong correlation $(p=0.000402)$ as reported in Wiggins et al. \cite{wiggins2022public}.

The scatter plot of average exam grade versus engagement rate is shown in Fig.~\ref{f:Performance}.  Because there were only 3 total exams, the only possible participation rates for a given student are $0\%, 33\%, 67\%$, and $100\%$; in our case, all students participated at least once in the revision process.  As written on the plot, the line of best fit has the equation $y = 0.142 x + 50.68$, indicating a positive correlation with strength $14.2\%$ between participation rate and average exam grade.

To determine whether the slope is significantly positive (a positive correlation of participation with exam grades), we studied the variation in the slope if each point were dropped, one at a time.  That is, for each data point $i$, we construct the line of best fit $y = m_{(i)} + x b_{(i)}$ with point $i$ removed.  Then we report the uncertainty in the slope as the standard deviation
\begin{align}
    \delta m = \sqrt{ \sum_{i = 1}^n 
    \frac{\big( m_{(i)} - \langle m \rangle \big)^2}{n-1} }
\end{align}
In this way, we obtain a slope and associated uncertainty
\begin{align*}
    m = 0.142 \pm 0.028   \: ,
\end{align*}
which suggests a statistically significant positive correlation between participation rate in the public exams revision process and exam grades.  We can further quantify the statistical significance of this correlation by computing a corresponding $p$ value.  If we interpret the central value and uncertainty $\langle m \rangle \pm \delta m$ as the mean and standard deviation of a normal distribution, then we can directly calculate the $p$ value from the one-sided integral
\begin{align}
p = \sqrt{\frac{1}{2\pi (\delta m)^2}} \int\limits_{-\infty}^0 dm \: e^{- \frac{(m - \langle m \rangle)^2}{2 (\delta m)^2}} = 3.1 \times 10^{-7} \: ,
\end{align}
which is highly statistically significant.  Thus, given the differences in the two studies -- the lower threshold for counting as ``participating'' in our study and measuring the correlation with exam average rather than overall grade -- we found a positive impact of participation which is at least as statistically significant as that observed by Wiggins et al. \cite{wiggins2022public}.

%
\subsubsection{Motivation}
%

We also studied quantitative data collected from online MSLQ surveys \cite{pintrich1991manual} which were collected during the first week of class. The MSLQ is based on the general cognitive view of motivation and learning strategies\cite{duncan1991manual}, with students rating themselves on a seven-point Likert scale from ``not at all true of me'' to ``very true of me,'' as described in Sec.~\ref{sec:Methods}.  In our study, we utilized only the motivation section of the MSLQ, which includes 18 items that assess students' goals, value beliefs for a course, and self-efficacy. The value components of motivational scale are intrinsic goal orientation (3 items), extrinsic goal orientation (4 items), task value (4 items) and the expectancy component is self-efficacy (7 items).

\begin{figure}[t]
    \centering
    \includegraphics[width=0.7\textwidth]{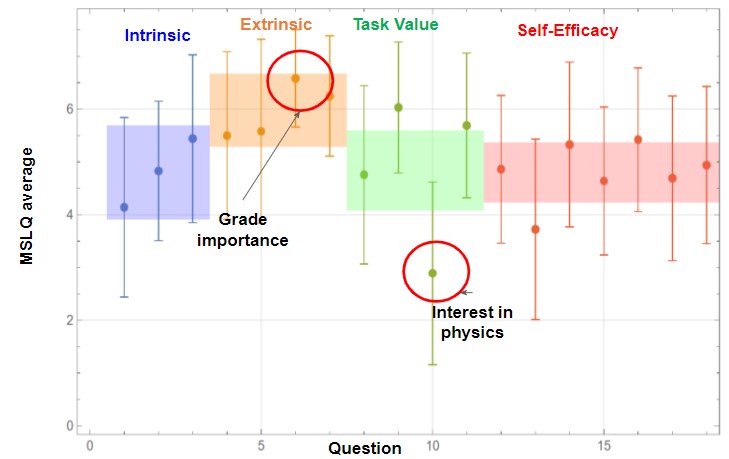}
    \caption{Correlation between the responses of students of different items in MSLQ surveys.  Colored bands show the mean and standard deviation for items measuring the same component.}
    \label{f:MSLQ_surveys}
\end{figure}

Fig.~\ref{f:MSLQ_surveys} shows the correlation between the responses of the students of items that fall on the same component. We found the average and standard deviation of MSLQ items for each component (intrinsic and extrinsic motivation, task value and self-efficacy). Students' responses on the MSLQ show modest perception of intrinsic value and higher levels of extrinsic motivation.  Students had particularly low interest in physics for its own sake but particularly high emphasis on grade importance.  Most items were consistent with a neutral response (4) or slightly positive on the Likert scale.  These observations can explain the observed participation rates seen in Fig.~\ref{f:Engagement}.  The rate of ``engagement'' which carried a grade incentive and a low barrier to entry was very high ($>80\%$ across all exams).  And while the number of questions edited was small (initially 2 on the first midterm), that rate grew over time and carried no additional grade incentive.  This suggests that students were incentivized to engage in at least minimal participation with the public exams revision process because of the grade points, but that genuine engagement (revising questions) also increased over time.  This could indicate that the public exam process helped students overcome their initial disinterest in physics to achieve more authentic engagement with the subject.

In the future, multiple questions about grade importance could be asked to increase the statistical significance of the data.  We could also ask about grade importance and extrinsic motivation in future open–ended written responses to get more detail.

We also compared the MSLQ scores for students who commented positively versus negatively about aspects of the Public Exams. The class average for MSLQ anxiety score was $5.82 \pm 1.66$. In the open-ended responses, 6 students commented positively about the impact of public exams on their anxiety; their average MSLQ anxiety score was $5.61 \pm 1.25$.  Only one student commented negatively about the impact of public exams on their anxiety, with an average MSLQ anxiety score of $6.00 \pm 0.82$.  Among students who did not comment about the impact of public exams on anxiety, the average MSLQ anxiety score was $5.86 \pm 1.72$.  There were no students who commented both positively and negatively about the impact of public exams on anxiety.  The differences in average MSLQ anxiety scores are not statistically significant between these groups, suggesting that the students who benefited from reduced test anxiety were typical of the overall class's anxiety levels.  

Regarding self efficacy, the MSLQ survey contains 7 relevant items, with a class average self-efficacy score of $4.80 \pm 1.59$.  In the open-ended responses, 2 students commented positively about the impact of public exams on their self confidence; their average MSLQ self-efficacy score was $3.14 \pm 1.64$ (around 1$\sigma$ below the class average).  These students gave their comments in response to the question, ``Is the second midterm exam (Public Exam) better than the first midterm exam.  Please explain.''  One student, whose MSLQ self-efficacy score was $1.6$ ($2\sigma$ below class average),  answered,  
\begin{center}
  \textit{
        ``Yes, I felt more confident going into midterm 2 and the final than I did going into midterm 1.``}.
 \end{center}
    Another student, whose MSLQ self-efficacy score was $4.7$ ($0.1\sigma$ below class average), answered
\begin{center}
        \textit{ ``Yes, because I wasn't confident in the public exam till I studied, But w/ more questions the actual test covered concepts I knew.``}
\end{center}
The interpretation of this question is somewhat unclear because of the phrasing, but it is noteworthy that the first student scored significantly lower in self-efficacy on the MSLQ survey, but then experienced a growth in confidence over time.  This could be indicative of positive benefits from the public exam system on self efficacy.  There were no students who made negative comments about the impact of public exams on self confidence.

%
\section{Discussion}
\label{sec:Discussion}
%

In assessing the overall impact of public exams, in our study we found that the comments related to public exams were on average \textit{less positive} than the control, with weak significance ($p = 0.14$). This is in direct contrast with the strongly significant increase in positive comment rate for public exams seen by Wiggins et al.  We also re-analyzed Wiggins et al.'s published data using our exact methodology and generally confirmed the conclusions of their analysis.  Regarding the impact of engagement on grades in our study, we saw a modest but statistically significant correlation $m = 0.142 \pm 0.028$ between the student participation rate and average exam grade.  This suggests that the correlation of exam performance with student engagement is positive, with a one-sided $p$ value of $p = 3.1 \times 10^{-7}$ assuming a normal (Gaussian) distribution.  This is consistent with the statistically significant correlation reported in Wiggins et al.

Beyond the overall positive or negative impact of public exams, we also studied various specific benefits of public exams reported by Wiggins et al.  These included directing to core concepts, reducing anxiety, overcoming language barriers, deepening thought, and promoting authentic engagement.  Wiggins et al. also hypothesized a possible benefit of enhanced collaboration but did not see evidence of it in their study.

Regarding the impact of public exams on directing students to core concepts, we found a significant number of both positive and negative student comments.  We performed a Pearson $\chi^2$ test for independence and a T test on the positivity rate of these comments versus the control. For this data, there was a significant student-by-student correlation between the positivity rate for directing to core concepts and the positivity rate of control ($p = 0.23$).  The positivity rate for Public Exams helping direct students to core concepts was $(84.4 \pm 30.1)\%$, whereas for control the positivity rate was $(98.0 \pm 5.5)\%$.  The decrease in positivity rate for Public Exams directing to core concepts compared to control was statistically significant: $p = 0.009$.  This shows that students reported a significantly reduced benefit in our study in physics, compared to that of Wiggins et al. in a biology context.

Regarding the impact of public exams on reducing anxiety, we found 7 students commented as discussed above.  A Pearson $\chi^2$ test for independence on this small sample was inconclusive, but the mean positivity rate for comments about Public Exams' impact on anxiety was $(85.7 \pm 37.8)\%$ while the positivity rate for control was $(98.0 \pm 5.5)\%$.  The reduction in positive comments about Public Exams' impact on anxiety compared to control was statistically significant ($p = 0.05$).

Regarding overcoming language barriers, deepening thought, promoting authentic engagement, and enhanced collaboration, our results were less conclusive.  We did not ask any specific questions regarding the impact on language barriers or enhanced collaboration, and none of the students' comments related to these issues in the open-ended responses.  We identified only a single student comment each relating to deepening thought and promoting authentic engagement: 
\begin{center}
\textit{
``I would prefer having more public questions because I learn these concepts much more deeply and feel more prepared to take the exam.``}.
\end{center}
\begin{center}
\textit{
``I was most engaged when we were going over practice problems and the public exam.``}.
\end{center}

Independent quantitative data from the Motivated Strategies for Learning Questionnaire (MSLQ) gave measures of student test anxiety and self-efficacy.  
The average MSLQ anxiety score for the class was $5.82 \pm 1.66$ on a 7-point Likert scale, and we found no significant differences in MSLQ anxiety score among students who said that public exams helped their anxiety, those who said public exams worsened their anxiety, and those who did not comment about test anxiety.  
Similarly, the average MSLQ self-efficacy score for the class was $4.80 \pm 1.59$, and two students made comments indicating that public exams had increased their self-confidence.  These two students had MSLQ self-efficacy scores of $3.14 \pm 1.64$.  Although only a $1\sigma$ difference from the class average, this may suggest that students with lower self-efficacy were significantly benefited by the public exam system.

\section {Conclusions and Outlook}
\label{sec:Concl}

Our mixed-methods design research shows evidence of similar benefits to student outcomes and student experiences, as in  Wiggins et al.\cite{wiggins2022public}.  We found evidence of benefits to student outcomes (Fig.~\ref{f:Performance}), an overall positive perception of public exams (Fig.~\ref{f:T_test}), and positive impacts on test anxiety (Fig.~\ref{f:Anxiety_T_Test}) and directing to core concepts (Fig.~\ref{f:Core_Concepts_T_Test}). 
However, we also see meaningful differences from the unambiguously positive impacts reported by Wiggins et al.  In our study, the perception of public exams, while positive, was significantly \textit{less} positive than our control sample (Fig.~\ref{f:T_test}).  The same is true for the perceived impact on test anxiety (Fig.~\ref{f:Anxiety_T_Test}) and directing to core concepts (Fig.~\ref{f:Core_Concepts_T_Test}).  This is the opposite of what was seen in Wiggins et al., where the perception of public exams, both in general and with respect to these specific impacts, was significantly more positive than in the control sample.  In our study, we also observed a minority of students with strong negative opinions about public exams (see sections \ref{sec:PEcomments} and \ref{sec:anxiety} for quotes).

We hypothesize that these differences between the results of our study and that of Wiggins et al. could be due to the application of public exams in a \textit{physics }context, as opposed to biology.  
Physics as a discipline places particular importance of the application of procedural knowledge to solve problems and make predictions, and this application of procedural knowledge appeared as a persistent theme across our mixed-methods analysis.
Some student quotes explicitly stated that the withholding of information during the public exams process sometimes made it harder to know how to apply the procedural knowledge, and some suggested that no information should be withheld at all.
Moreover, students consistently identified the portions of the class involving applying procedural knowledge as the most helpful or engaging elements of the class (Fig.~\ref{f:helpful}) and recommended adding more activities that practice the application of procedural knowledge (Fig.~\ref{f:Suggestions}).
These data suggest that the withholding of information during the public exams process may have had a negative interaction with questions involving the application of procedural knowledge, which are centrally important in physics courses.

\begin{figure}[t]
    \textbf{Exam Question:} \qquad
    \begin{center}
    A cube made of crown glass $(n = 1.52)$ is submerged in water $(n = 1.33)$.  If a ray of light enters the cube with an angle of incidence in the water of $34.0^\circ$ determine the angle of refraction in the glass.
    
    \vspace{0.5cm}
    
    \begin{itemize}
        \item   a.) \qquad $25.34^\circ$
        \item   b.) \qquad $29.29^\circ$
        \item   c.) \qquad $39.72^\circ$
        \item   d.) \qquad $41.25^\circ$
    \end{itemize}
    \end{center}

    \vspace{0.25cm}

    \textbf{Public Question:} \qquad 
    \begin{center}
    A crown glass cube $(n = 1.52)$ is submerged in water $(n = 1.33)$.  If a ray of light enters the cube with an angle of incidence in the water of [ANGLE WITHHELD] determine the angle of refraction in the glass.
    
    \vspace{0.5cm}
    
    \begin{itemize}
        \item   a.) \qquad [OPTION WITHHELD]
        \item   b.) \qquad [OPTION WITHHELD]
        \item   c.) \qquad [OPTION WITHHELD]
        \item   d.) \qquad [OPTION WITHHELD]
    \end{itemize}
    \end{center}

    \vspace{0.25cm}

    \textbf{Alternative Question:} \qquad 
    \begin{center}
    A transparent cube of an unknown substance is submerged in water $(n = 1.33)$.  If a ray of light enters the cube with an angle of incidence in the water of $25^\circ$ and the angle of refraction in the cube is $18^\circ$, determine the refractive index $n$ of the cube.

    \vspace{0.5cm}
    
    \begin{itemize}
        \item   a.) \qquad 1.00
        \item   b.) \qquad 1.33
        \item   c.) \qquad 1.82
        \item   d.) \qquad 1.95
        \vspace{0.25cm}
        \item   \textit{Ans:  c}
    \end{itemize}
    \end{center}
    
    \caption{An alternative to withholding information in the public questions which still tests the same procedural knowledge.}
    \label{f:MarahattaAlternative}
\end{figure}

Our interpretation is therefore that the public exam system should be modified to better adapt it to a physics context.  Instead of withholding parts of the questions, which confused some students in this study, physics-specific public exams should reveal 100\% of the information in a question and make use of the ability to change the values in a problem or rearrange the same calculation in a different way.  Consider the example shown in Fig.~\ref{f:MarahattaAlternative}.  This question deals with Snell's law, which states the relationship between the angle of incidence $\theta_i$, angle of refraction $\theta_r$, and the refractive indices $n_i, n_r$ of the medium of incidence and medium of refraction:
\begin{align}   \label{e:Snell}
    n_i \, \sin\theta_i = n_r \, \sin\theta_r   \: .
\end{align}
On the final exam (top panel of Fig.~\ref{f:MarahattaAlternative}), students are given $n_i$, $n_r$, and $\theta_i$, and asked to recall Snell's law and solve it for $\theta_r$.  The corresponding public question on the public final exam (middle panel of Fig.~\ref{f:MarahattaAlternative}) maintained an identical structure to the intended final exam question, but provided only $n_i$ and $n_r$, withholding both $\theta_i$ and possible options for $\theta_r$.  A public question like this is potentially confusing for students, because it does not give sufficient information to actually \textit{apply the procedure} of solving Snell's law, nor to determine whether a student has applied it \textit{correctly}.  

Instead, we propose to generate an alternative question, such as the one shown in the bottom panel of Fig.~\ref{f:MarahattaAlternative}.  This question is worded differently from the other two, but it still tests the same procedure: applying Snell's law \eqref{e:Snell}.  This alternative question reveals 100\% of the relevant information, which allows students to actually \textit{perform the calculation} of Eq.~\eqref{e:Snell}; moreover, providing the correct answer to the alternative question allows students to confirm that they have applied the procedure \textit{correctly}.  Despite giving away 100\% of the information on the alternative question, the intended exam question is uncompromised, because it tests the same procedure \eqref{e:Snell} in a different way.  Note that the alternative question does not simply change the numerical values in the problem;
this ensures that students understand how to rearrange and use the equations instead of memorizing the procedure shown on the alternative question.  Furthermore, by providing the full information in the alternative version,  students can focus directly on applying the core procedure, saving time and the frustration of trying to guess the withheld information.

An IRB-approved study of these physics-specific public exams and their classroom impact would be a natural follow-up project to this one.  In such a future study, we of course recommend expanding the data sample to cover more students, sections, and Instructors to maximize the statistical reach of the analysis.  We recommend keeping the mixed-methods approach employed here, but with a few changes to study methodology.  We recommend that a future study include a dedicated control group of students, if possible -- that is, students taking the same course under a traditional exam format.  We also recommend exploring policies to increase student motivation and participation during the editing and revision of questions during the public exams process.  Possibilities could include a small extra credit incentive or a group practice session in which students edit poorly written questions and improve their clarity as a group.  Future studies should expand on the open-ended survey questions used to collect qualitative data about student experiences with public exams.  To maximize the impact of the study, these open-ended surveys should focus more on the impact of public exams on student outcomes and less on the overall outcomes of the class.  Adding specific questions relating to its possible effects on language barriers, deepening thought, promoting authentic engagement, or enhancing collaboration could be useful as well.  We also recommend conducting multiple iterations of the MSLQ surveys to gauge student responses over time.  Changes to the data analysis in the future can also include a comparison of student performance on publicly-revealed questions versus hidden questions to further test the impact of public exams.

In this work, we have tested the transferability of a novel assessment system from a biology context to a physics context. Our results indicate that this innovative approach to assessment \textit{can} be beneficially applied in this context, but likely \textit{needs modification} to optimize it for the cognitive tasks specific to physics, particularly the application of procedural knowledge. This work contributes to expanding the reach of innovations in assessment within STEM education.


\section*{Acknowledgments}

\label{sec:acknowledgment}

BM and MS acknowledge Dr. Michael Paolone and Dr. Krishna Acharya (instructor of the introductory algebra-based physics course) for adapting the Public Exam system in the Department of Physics at New Mexico State University for the first time and for collecting data.  BM and MS also thank Dr. Benjamin Wiggins for a critical reading of the manuscript and constructive feedback.


\section*{Data Availability}

\label{sec:data}

Raw individualized, anonymized data used in this study, including open-ended responses,  quantitative grade data, and MSLQ survey data, are available in a public repository \cite{marahatta_2026_22711553}.

%

%


\end{document}